\documentclass[10pt,journal,final,letterpaper,twocolumn,twoside]{IEEEtran}

\usepackage{amsmath,amsfonts,bm,amssymb,psfrag,ifthen,color,subfigure}
\usepackage{mathrsfs}
\usepackage[justification=centering,font=footnotesize]{caption}
\usepackage[dvips]{epsfig}
\usepackage[dvips]{graphicx}
\usepackage{amsmath,amsfonts,bm,amssymb,psfrag,ifthen,color,subfigure}
\usepackage{cases}
\usepackage{algpseudocode}
\usepackage{algorithm}
\usepackage{algorithmicx}
\usepackage{ifthen}
\usepackage{setspace}
\usepackage{cite}
\usepackage{url}
\usepackage{longtable}
\usepackage{stfloats} 
\usepackage{graphics,booktabs,color,subfigure}
\usepackage{float}

\usepackage{listings}

\allowdisplaybreaks[4]

\begin{document}
\title{Robust Power Allocation for Integrated Visible Light Positioning and Communication Networks}

\author{Shuai Ma,  Ruixin Yang, Chun Du, Hang Li, Youlong Wu, Naofal Al-Dhahir, and Shiyin Li

\thanks{S. Ma is with the Peng Cheng Laboratory, Shenzhen 518055, China. (e-mail: mash01@pcl.ac.cn)}
%
%
%
%
}

\maketitle
\begin{abstract}

Integrated visible light positioning and communication (VLPC), capable of combining advantages of visible light communications (VLC) and visible light positioning (VLP), is a promising key technology for the future Internet of Things.
In VLPC networks, positioning and communications are inherently coupled, which has not been sufficiently explored in the literature.
We propose a robust power allocation scheme for integrated VLPC Networks by exploiting the intrinsic relationship between positioning and communications.
Specifically, we derive explicit relationships between random positioning errors, following both a Gaussian distribution and an arbitrary distribution, and channel state information errors.
Then, we minimize the Cramer-Rao lower bound (CRLB) of positioning errors, subject to the rate outage constraint and the power constraints, which is a chance-constrained optimization problem and generally computationally intractable.
To circumvent the nonconvex challenge, we conservatively transform the chance constraints to deterministic forms by using the Bernstein-type inequality and the conditional value-at-risk for the Gaussian and arbitrary distributed positioning errors, respectively, and then approximate them as convex semidefinite programs.
Finally, simulation results verify the robustness and effectiveness of our proposed integrated VLPC design schemes.

\end{abstract}

\begin{IEEEkeywords}
Visible light communication, Visible light positioning, Robust power allocation, Outage probability.
\end{IEEEkeywords}

\section{INTRODUCTION}\label{sec:introduction}

As predicted by Cisco, over  $500 $ billion Internet of Things (IoT) devices will be connected to the Internet globally by 2030, and about $80\%$ of the mobile traffic occurs in indoor environments\cite{CISCO_2012}.
This would also lead to growing demands for indoor location-based services, which are supported by accurate positioning and high data rates\cite{Farahsari2022,Alam2021}.
The related applications include smart manufacturing, safety monitoring, logistics management, and indoor navigation.
For indoor environments, the global navigation satellite system (GNSS) signals are usually weak due to the multipath effect and signal blocking. Meanwhile, conventional radio frequency (RF) based indoor positioning approaches, e.g., Wi-Fi, radio frequency identification (RFID), and ultra-wideband (UWB), generally suffer from electromagnetic interference, sensitivity to the changing environment, and multipath effects.
Besides, the RF spectrum scarcity is also a challenging issue as the number of electronic devices sharing it becomes large.

Visible light positioning and communications (VLPC) provide alternative solutions for indoor location-based services. At present, most of the existing works study visible light positioning (VLP) and visible light communications (VLC) separately\cite{Kamalakis2022,MemediFirstquarter2021,MatheusFourthquarter2019,Farahsari2022,Alam2021}. Some recent advances in integrated VLPC networks motivate us to investigate a unified system, capable of fulfilling the requirements of both positioning and communications \cite{Luo_survey_2017,Keskin}. Specifically, in the integrated VLPC network, both the positioning and communication operations can be jointly optimized via the shared use of a single hardware platform and a joint signal processing framework.
Owing to its inherent advantages, including, but not limited to, no electromagnetic interference, low multipath effects, and low
deployment cost, integrated VLPC networks have drawn more and more research attention in both industrial and academic communities.

For VLP, there are many positioning algorithms based on received signal strength (RSS) \cite{Zhang2014,ALAM_ACCESS_2019,DU_ACCESS_2019,Sun2022}, hybrid RSS/angle of arrival (AOA) \cite{Sahin_JLT_2015}, time of arrival (TOA)\cite{Sun_2015_Dec}, time difference of arrival (TDOA) \cite{Gu_2016_May,Du_IPJ_2018} and the maximum likelihood estimators \cite{Furkan_TCOM_2018}.
In \cite{Lin_CL_2020},  a position estimation deep neural network (PE-DNN) aided VLC receiver was designed to achieve data transmission and location positioning simultaneously.
In \cite{Lin_IPJ_2017}, the authors experimentally demonstrated an indoor VLC and a VLP system using the orthogonal frequency division multiple access (OFDMA).
The integrated VLPC system also was proposed by other different modulations or signal processing, such as code division multiple access (CDMA)\cite{Chen_OE_2022_Integrated}, $m$-CAP\cite{Shi_TBC_2023_joint}, retroreflectors\cite{Shao_JIOT_2023}, and adaptive feedback threshold\cite{Jin_OE_2022_Adaptive}.
Based on the reinforcement learning framework, an intelligent resource allocation scheme was studied in \cite{Yang_WCL_2019} for integrated VLC and VLP systems, where the sum rate is maximized subject to the constraints of minimum data rates and positioning accuracy.
In \cite{XU_2017_OE}, an OFDMA-based integrated VLPC system was proposed to estimate the receiver's position based on the power of the data sequence.
By using filter bank multicarrier-based subcarrier multiplexing (FBMC-SCM) and phase difference of arrival, the authors in \cite{Yang_PTL_2018} experimentally tested the integrated VLPC system.
Based on FBMC-SCM, a joint subcarrier and power allocation method was presented in \cite{Yang_TWC_2020} for multi-cell integrated VLPC systems to maximize the sum rate under both the minimum data rate and positioning accuracy constraints.
For the integrated VLPC IoT network \cite{Yang_IOT_2020}, the authors jointly optimized AP selection, bandwidth allocation, adaptive modulation, and power allocation to maximize the data rate.
Note that, in the above existing literature, the estimated position information is not fully utilized in the data transmission.
In \cite{Pal_TWC_2022_Channel}, the relationship between channel gain distributions and UE's position and orientation was studied, and the symbol error probability was derived based on the least-square channel estimation and channel gain distributions, while the relationship between positioning and communication performance is not revealed.
In \cite{Ma_TCOMM_2022_optimal}, the intrinsic relationship between positioning errors and channel estimation errors was derived, and a robust joint power allocation scheme was proposed through the time division multiple access (TDMA). However, this relationship is statistical and implicit.

In this paper, we study the integrated VLPC network based on the frequency division multiple access (FDMA) to transmit VLP and VLC signals simultaneously. Then, we reveal the fundamental relationship between positioning and communication performance, in terms of exploring the intrinsic explicit relationship between positioning errors and channel estimation errors. Moreover, we propose two robust power allocation schemes by addressing Gaussian distributed and arbitrary distributed positioning errors, respectively.

Accordingly, the contributions of this paper are summarized as follows:

\begin{itemize}
\item Based on the derived the Cramer-Rao lower bound (CRLB) of VLP and achievable rates of VLC,  we describe the relationship between positioning errors and channel estimation errors for the integrated VLPC network, and there exists an optimal tradeoff relationship between the VLP and VLC. On the one hand, VLP can enhance VLC, i.e., location information can be used to increase the information transmission signal-to-noise ratio (SNR) and rate. On the other hand, VLP and VLC are mutually restricted. Both positioning accuracy and communication rate depend on the allocated power.
Moreover, positioning accuracy affects channel estimation error, and channel estimation error further affects the communication rate.
To our best knowledge, this inherent coupling between VLP and VLC is revealed for the first time.

\item Next, we propose a robust power allocation optimization framework to minimize the CRLB of VLP subject to the rate outage chance  constraint and the optical and electrical power constraint.
This chance-constrained optimization problem is generally intractable. Meanwhile, due to the matrix inverse, the problem usually is nonconvex, which make the problem more challenging.
\begin{itemize}
    \item For Gaussian distributed positioning errors, we conservatively transform the chance  constraint to a deterministic form, based on the Bernstein-type inequality to circumvent the intractability of the chance  constraints.
    Furthermore, the non-convex deterministic form constraints are  approximated by a convex form based on the matrix norm feature.
    \item Arbitrary distributed positioning errors are a more practical scenario to cover Gaussian and non-Gaussian distributions, where only the mean and variance are known, but the distribution form is uncertain.
    To tackle a variety of the chance constraint on the uncertainty set, the worst-case distribution of the Conditional Value-at-Risk (CVaR) is conservatively approximated to a more tractable form.
    Then, by adopting successive convex approximation (SCA), the joint  nonconvex optimization problem can be transformed into a series of convex subproblems.
\end{itemize}
Finally, both of these  probability-constrained problems can be reformulated as a convex semidefinite program (SDP), which can be solved by the interior point method.
\end{itemize}

The rest of this paper is organized as follows.
We present the VLPC system model in Section \ref{sec:sys_model}.
The key performance metrics for the VLPC system are derived in Section \ref{sec:tradeoff-between-positioning-and-communication}.
In Section \ref{sec:robust-integrated-vlpc-design}, we investigate the chance-constrained robust integrated VLPC design.
Extensive simulation results are presented in  Section \ref{sec:numerical-results}. Section \ref{sec:conclusion} concludes the paper. Moreover, Tables \ref{Tab:acronyms} and \ref{Tab:notations}   present  the  main acronyms and the key notations of  this paper, respectively.

\emph{Notations}: Boldfaced lowercase and uppercase letters represent vectors and matrices, respectively.
$\mathcal{M}\triangleq \left\{1, \ldots ,M\right\}$. $\mathrm{Tr}\left(\cdot\right)$ and $\left(\cdot\right)^T$ denote the trace and transpose of a matrix, respectively.
$\mathbf{I}$ denotes the identity matrix. $\mathbb{S}^{n}$ represents the space of  $n $-dimensional real symmetric matrices.
$\mathbb{R}^{n}$ represents the space of  $n $-dimensional real vectors. $\mathbf{0}$ is a column vector where all elements are  $0 $.
\(\bigl\|\cdot\bigr\|\) denotes the norm of a vector or the \(2\)-norm of a matrix.
\(\bigl\|\cdot\bigr\|_{\text{F}}\) denotes the Frobenius norm of a matrix.

\begin{table}[htpb]
    \caption{Summary of Main Acronyms}
    \label{Tab:acronyms}
    \centering
    \begin{tabular}{|c|l|}
        \hline
        Notation & Description\\
        \hline
        VLP & Visible light positioning\\
        \hline
        VLC & Visible light communication\\
        \hline
        VLPC & Visible light positioning and communication\\
        \hline
        UE & User equipment\\
        \hline
        RSS & Received signal strength\\
        \hline
        CRLB & Cramer-Rao lower bound\\
        \hline
        CSI & Channel state information\\
        \hline
        SNR & Signal-to-noise ratio\\
        \hline
        SDP & Semidefinite program\\
        \hline
        CVaR & Conditional Value-at-Risk\\
        \hline
        SCA & Successive convex approximation\\
        \hline
    \end{tabular}
\end{table}

\begin{table}[htpb]
    \caption{Summary of Key Notations}
    \label{Tab:notations}
    \centering
    \begin{tabular}{|c|l|}
        \hline
        Notation & Description\\
        \hline
        $\mathbf{p}_{\text{p}}$ & Allocated positioning power vector\\
        \hline
        $ P_{\text{p},i}$ & Allocated positioning power for $i$-th LED\\
        \hline
        $ P_{\text{c}}$ & Allocated communication power\\
        \hline
        $ P_{\text{total}}$ & Maximum total power\\
        \hline
        $ \mathbf{u}$ & Location vector of UE\\
        \hline
        $ \hat{\mathbf{u}}$ & Estimated location vector of UE\\
        \hline
        $ \mathbf{e}_{\text{p}}$ & Positioning error vector\\
        \hline
        $ \mathbf{v}_i$ & Location vector of $i$-th LED\\
        \hline
        $g_i$ & CSI between $i$-th LED and UE\\
        \hline
        $ \hat{g}_i$ & Estimated CSI between $i$-th LED and UE\\
        \hline
        $\Delta g_i$ & CSI estimation error between $i$-th LED and UE\\
        \hline
        $i^*$ & Index of the data-transmission LED  \\
        \hline
        $T_{\text{p}}$ & Duration of the positioning subframe\\
        \hline
        $ \mathbf{J}_{\mathbf{u}}\left(\mathbf{p}_{\text{p}}\right)$ & Fisher information matrix\\
        \hline
        $ R_{\text{c}}$, $ R_{\text{L}}$ & VLC achievable rate and its lower bound \\
        \hline
        $\mathcal{P}$ & Set of distributions for ${\Delta {g}}_{i^*}$\\
        \hline
        $ \bar{r}$ & Minimum rate requirement\\
        \hline
        $ P_{\text{out}}$ & Maximum tolerable outage probability\\
        \hline
        $\mathcal{M}$ & Set of the LED's index\\
        \hline
    \end{tabular}
\end{table}

\section{SYSTEM MODEL}\label{sec:sys_model}

\begin{figure}[htbp]
    \centering
    \begin{minipage}[b]{0.3\textwidth}
        \centering
        \includegraphics[width =\textwidth]{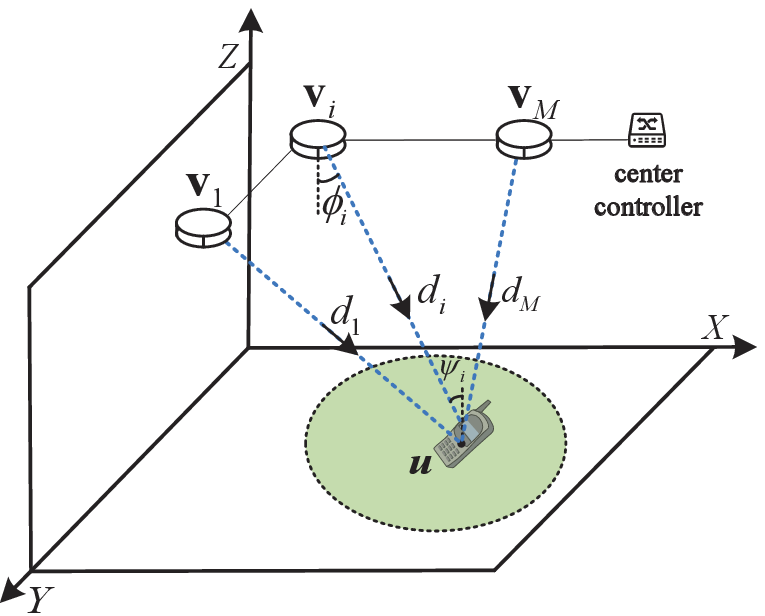}
        \vskip-0.2cm\centering {\footnotesize (a)}
    \end{minipage}
    \begin{minipage}[b]{0.5\textwidth}
        \centering
        \includegraphics[width =0.9\textwidth]{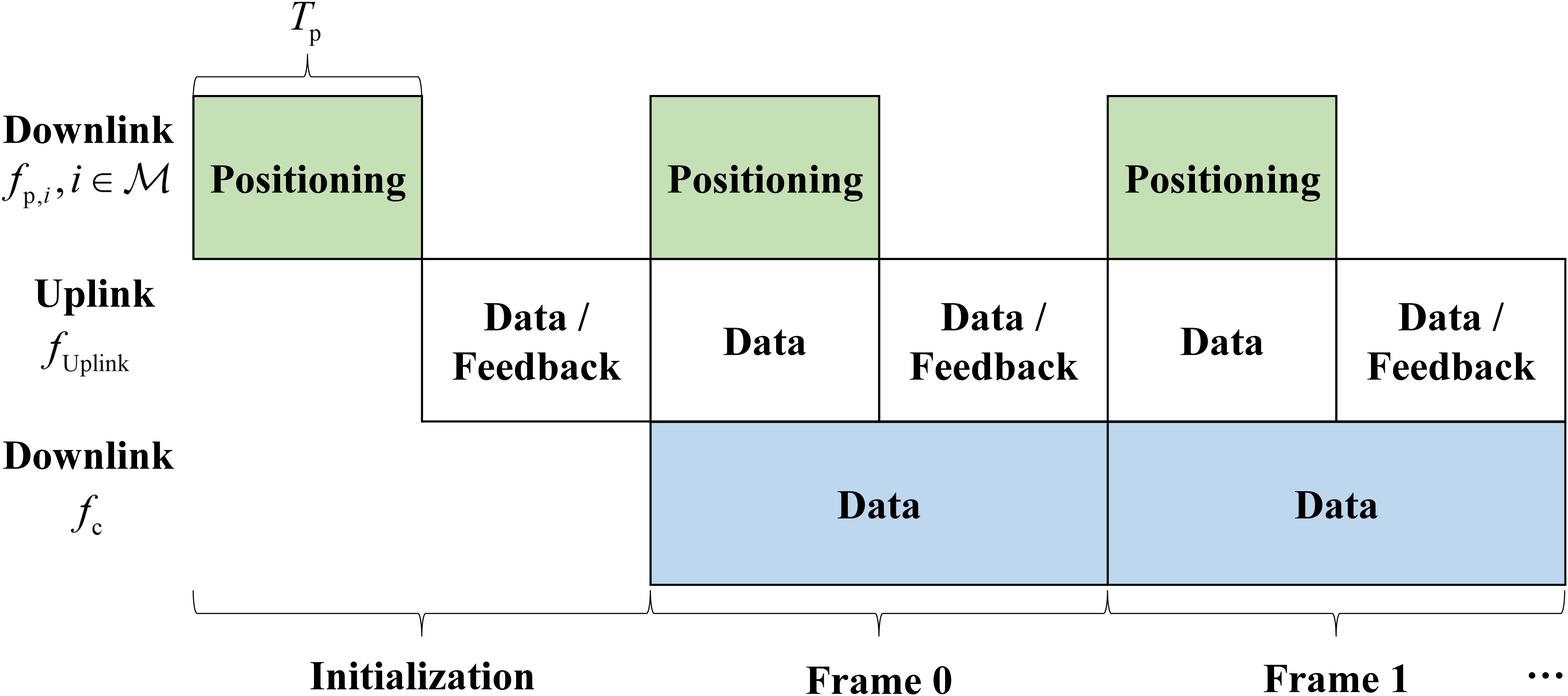}
        \vskip-0.2cm\centering {\footnotesize (b)}
    \end{minipage}
    \caption{The System model (a) ~The diagram of an  integrated VLPC network; (b) ~Frame structure of VLPC system.}
    \label{system}\label{frame_struct}
\end{figure}

As illustrated in Fig. \ref{system}(a), consider an  integrated VLPC network that includes a central controller,  $M\geq 3$  non-collinear LEDs, photodetector-based user equipment (UE), such as robots and automated guided vehicles
move on the factory ground, where $\mathbf{u} = \left[x_{\mathrm{u}},y_{\mathrm{u}},z_{\mathrm{u}}\right]^T$ and $\mathbf{v}_i = \left[x_{i},y_{i},z_{i}\right]^T$ denote the locations  of the UE and  the $i$-th LED ($i \in \mathcal{M}$), respectively.
The center controller connects all LEDs and controls the transmit signal of each LED. It also can tackle and share information from all LEDs and uplinks.
Without loss of generality, characteristics of the LEDs and the UE's photodetector (PD) can be known by  measurement or sensors. To simplify the analysis, we assume that the $M$ LEDs are identical, the orientation of the UE's  PD is upward, all LED orientation is downward, and all LED positions are exactly known.

To avoid interference among different functions, the communication frame of the system  includes a downlink positioning subframe, a downlink data subframe, and an uplink subframe  with different frequencies $f_{\mathrm{p},i},i\in\mathcal{M}$, $f_{\mathrm{c}}$ and $f_{\mathrm{Uplink}}$, respectively, as shown in  Fig. \ref{frame_struct}(b).
This structure can be implemented by the classical frequency-division multiple access. Besides, the allocated frequencies $f_{\mathrm{p},i}$ is an ID of $i$-th LED to distinguish the source of the positioning signals.
For the positioning subframe with the duration $T_{\mathrm{p}}$, $M$ LEDs simultaneously transmit positioning signals to  the UE, which can estimate the RSS.
Although the center controller can use the RSS information to calculate the UE's location, the issues of the UE privacy and the computing load at the center controller make it reasonable that the UE directly undertakes all the computation tasks.
Thus, we assume that the UE is powerful enough to perform the positioning, channel state information (CSI) estimation, and the joint power allocation for the next subframe.
In  the uplink feedback subframe, the UE only feeds back the CSI and the result of the power allocation to the center controller by infrared communications or RF communications.
Finally, the  center controller deploys the power allocation results, and selects the LED with the best CSI to transmit data to the UE  during  the next downlink data subframe. The model can be extended into the multi-UE scenario through a proper multiple-access protocol, which can refer to the classical theory of multi-user networks. For example, based on the TDMA, each UE will be independently served at the allocated slot, which is equivalent to the single-UE case.


\subsection{Downlink Positioning Subframe}
In the positioning subframe,  the UE  estimates the RSS based on the received signals from LEDs.
Specifically, for $t\in\left[ {0,{T_{\rm{p}}}} \right]$, the  $i$-th LED transmits the positioning  symbol ${{s_{{\mathrm{p}},i}}\left( t \right)}$,  where $ - A \le {{s_{{\mathrm{p}},i}}\left( t \right)} \le A$, $\mathbb{E}\left\{ {{{s_{{\mathrm{p}},i}}\left( t \right)}} \right\} = 0$, $\mathbb{E}\left\{ {s_{{\mathrm{p}},i}^2\left( t \right)} \right\} = \varepsilon  $,   $A > 0$ and  $\varepsilon  > 0$.
For brevity,  we   drop the time index $t$ throughout the paper, where ${{s_{{\mathrm{p}},i}}\left( t \right)}$ is denoted by ${{s_{{\mathrm{p}},i}}}$.
Hence, the transmitted positioning signal ${{x_{{\mathrm{p}},i}}}$ of the $i$-th LED is given as
\begin{align}\label{x_p}
x_{{\mathrm{p}},i} = \sqrt{P_{\mathrm{p},i}} s_{\mathrm{p},i} + {I_{{\mathrm{DC}}}},\forall i \in \mathcal{M},
\end{align}
where ${{P_{{\mathrm{p}},i}}}$ indicates the allocated positioning power of the $i$-th LED, and ${I_{{\rm{DC}}}}>0$ denotes the direct current (DC) bias.

At the  UE side, the  received light is detected by the PD and usually travels via the line of sight (LOS) and diffuse links, and a DC blocking circuit is adopted to filter the DC component. The received positioning signal $y_{{\mathrm{p}},i}$ from the $i$-th LED  can be expressed as
\begin{align}\label{y_p}
{y_{{\mathrm{p}},i}}  = {g_i}\sqrt{P_{\mathrm{p},i}} s_{\mathrm{p},i}  + {n_{{\mathrm{p}},i}},\forall i \in \mathcal{M},
\end{align}
where $g_i$ is the channel gain between the $i$-th LED and the UE, and ${n_{{\mathrm{p}},i}}$ is the zero-mean Gaussian noise $\mathcal{N}\left( {0, \sigma _{\mathrm{p}}^2} \right)$.

Generally, the channel gain $g_i$ can be decomposed as
\begin{align}
g_i \triangleq g_{i,\mathrm{LOS}} + g_{i,\mathrm{diffuse}}, i\in\mathcal{M},
\end{align}
where $g_{i,\mathrm{LOS}}$ and $g_{i,\mathrm{diffuse}}$ are the channel gains of the LOS and diffuse links between the $i$-th LED and the UE, respectively. For the LOS link, the channel gain $g_{i,\mathrm{LOS}}$ is given by\cite{Kahn}
\begin{align} \label{channel_gain}
g_{i,\mathrm{LOS}} =
\begin{cases}
    \frac{\left( m+1 \right) \eta _{\mathrm{c}}\eta _{\ell}A_{\mathrm{R}}}{2\pi d_{i}^{2}}\cos ^m\left( \phi _i \right) \cos \left( \psi _i \right), & \left| \psi _i \right|\le \psi _{\mathrm{FoV}};\\
    0,&\mathrm{otherwise},
\end{cases}
\end{align}
where $m \triangleq \left.-\ln 2\middle/\ln \left(\cos\left(\phi_{\frac{1}{2}}\right)\right)\right.$ is the Lambertian index of the LED, $\phi _{\frac{1}{2}}$ denotes the semi-angle,
$A_{\mathrm{R}}$ is the effective area of PD,
$\eta _{\mathrm{c}}$ is the  electric-optical conversion coefficient,
$\eta _{\ell}$ is the optical-electric conversion coefficient,
${\psi _{{\rm{FoV}}}}$ indicates the FoV of PD, ${d_i}$ denotes   the distance from the $i$-th LED to the UE,
${\phi _i}$ denotes the angle of irradiance of the $i$-th LED, and  ${\psi _i}$ denotes the angle of the PD incidence from the $i$-th LED.

On the other hand, the CSI of the diffuse links  is always unknown, and the gain of the LOS link is significantly higher than that of the diffuse link \cite{Fath_2013,Q.Wang_2013}.
Therefore, to simplify the analysis, we assume that the diffuse links can be ignored, and only the LOS link needs to be considered, i.e., $g_i=g_{i,\mathrm{LOS}}$.

Finally, according to the frequency components of the received positioning signal, the UE can detect which LED is in the UE's FoV and estimate the RSS.
Let $\tilde{\mathcal{M}}\subset \mathcal{M}$ denote the set of the received LED's index, and $\mathbf{p}_{y}\left(\mathbf{u},\left\{\mathbf{v}_i\right\}_{i\in\tilde{\mathcal{M}}}\right) \triangleq\left[p_{y,i},...\right]^T\in\mathbb{R}^{\tilde{M}}$ denote the  power  vector of the  received signals $\left\{y_{\mathrm{p},i}\right\}_{{i\in\tilde{\mathcal{M}}}}$ from the $\tilde{M}$ received LEDs to the UE, where $p_{y,i}=\mathbb{E}\left\{y_{\mathrm{p},i}^2\right\}$.
Without loss of generality, we assume that $P_{\mathrm{p},i}=0,i\notin\tilde{\mathcal{M}}$.

\subsection{Positioning and Channel Estimation}

When at least 3 non-collinear LEDs can be received, using the RSS-based positioning method, the UE's location estimation can be transformed into a RSS-related estimation problem. For example, utilizing the least squares method\cite{Cengiz2021}, the positioning result of the UE can be solved by $\hat{\mathbf{u}}=\arg\min_{\hat{\mathbf{u}}}\left\|\mathbf{p}_{y}\left(\mathbf{u},\left\{\mathbf{v}_i\right\}_{i\in\tilde{\mathcal{M}}}\right)-\mathbf{p}_{\hat{y}}\left(\hat{\mathbf{u}},\left\{\mathbf{v}_i\right\}_{i\in\tilde{\mathcal{M}}}\right)\right\|^2,$
where $\mathbf{p}_{\hat{y}}\left(\hat{\mathbf{u}},\left\{\mathbf{v}_i\right\}_{i\in\tilde{\mathcal{M}}}\right)$ denotes the corresponding received power vector at the UE's Location $\hat{\mathbf{u}}\in\mathbb{R}^{3}$.
Thus, the positioning error  ${{\mathbf{e}}_{\mathrm{p}}}$  is defined  as
\begin{align}\label{e_p}
{{\bf{e}}_{\rm{p}}} = {\bf{u}} - {\bf{\hat u}}.
\end{align}

Generally, the positioning error ${{\mathbf{e}}_{\mathrm{p}}}$ is assumed to follow two typical probability distribution models: Gaussian distribution  and arbitrary distribution.
\begin{itemize}
\item  \emph{Gaussian distribution}: If only the mean and variance are known, the Gaussian distribution can maximize the entropy of the uncertain parameter without additive constraints.
This model has been adopted in \cite{Wang2009,Lottici_JSAC_2002,Gezici_SPM_2005} because the CRLB is asymptotically achieved  by the maximum-likelihood (ML) estimator even for finite data size\cite{Kay_1993,Cheung_TSP_2004}.

\item  \emph{Arbitrary distribution}:  The location estimators
based on approximate ML\cite{Chan_TVT_2006,Cheung_ICASSP_2004}, and multidimensional
scaling (MDS) \cite{Wei_2008} can also achieve the  CRLB. However, the estimation errors do not
necessarily follow a Gaussian distribution\cite{Gustafsson_2005,Wang2009}.
In this case,  although the distribution of the position error ${{\bf{e}}_{\rm{p}}}$ is unavailable,
its mean and variance also can be assumed to be known {\cite{Torrieri_1984,Wang2009,Wei_2008}}.

\end{itemize}

Based on the RSS $\mathbf{p}_{y}\left(\mathbf{u},\left\{\mathbf{v}_i\right\}_{i\in\tilde{\mathcal{M}}}\right)$ and the signal $\left\{x_{\mathrm{p},i}\right\}_{i\in\tilde{\mathcal{M}}}$, we can select the $i^*$-th LED to transmit the downlink data, which is assumed to correspond to the maximum channel gain, i.e.,
\begin{align}
    i^* &= \arg\max_{i\in{\tilde{\mathcal{M}}}}\frac{\int_{0}^{T_{\mathrm{p}}} y_{\mathrm{p},i}\left(t\right)s_{\mathrm{p},i}\left(t\right) \,\mathrm{d}t }{\sqrt{P_{\mathrm{p},i}}\int_{0}^{T_{\mathrm{p}}} s_{\mathrm{p},i}^2\left(t\right) \,\mathrm{d}t}\nonumber\\
    &=\arg\max_{i\in{\tilde{\mathcal{M}}}}g_i+\frac{\int_{0}^{T_{\mathrm{p}}} n_{\mathrm{p},i}\left(t\right)s_{\mathrm{p},i}\left(t\right) \,\mathrm{d}t}{\sqrt{P_{\mathrm{p},i}}\varepsilon T_{\mathrm{p}}},
\end{align}
and the corresponding LED location is ${{\mathbf{v}}_{{i^*}}} = {\left[ {{x_{{i^*}}},{y_{{i^*}}},{z_{{i^*}}}} \right]^T}$.

If only the LOS link is considered, according to the channel model  \eqref{channel_gain}, the estimated CSI $\hat{g}_{i^*}$ between the  ${i^*}$-th LED and the UE is given by
\begin{align}\label{channel_gain3}
{{\hat g}}_{i^*} = \frac{{\left( {m + 1} \right){\eta _{\mathrm{c}}}{\eta _{\mathrm{\ell}}}{A_{\mathrm{R}}}{{\left( {{z_{{i^*}}} - {{\hat z}_{\mathrm{u}}}} \right)}^{m + 1}}}}
{{2\pi {{\left\| {{\mathbf{\hat u}} - {{\mathbf{v}}_{{i^*}}}} \right\|}^{m + 3}}}}= \frac{{\mu {{\left( {{z_{{i^*}}} - {{\hat z}_{\mathrm{u}}}} \right)}^{m + 1}}}}
{{{{\left\| {{\mathbf{\hat u}} - {{\mathbf{v}}_{{i^*}}}} \right\|}^{m + 3}}}},
\end{align}
where $\mu  = \frac{{\left( {m + 1} \right){A_{\mathrm{R}}}{\eta _{\mathrm{c}}}{\eta _\mathrm{\ell}}}}{{2\pi }}$.

Due to the positioning error ${{\bf{e}}_{\rm{p}}}$,  the channel estimation \eqref{channel_gain3} is imperfect. The corresponding estimated CSI error $\Delta {g}_{i^*}$ between the  ${i^*}$-th LED and the UE is defined as follows
 \begin{align}\label{delta_g}
     \Delta {g}_{i^*} ={{g}}_{i^*}-{{\hat g}}_{i^*}.
 \end{align}
Based on  \eqref{channel_gain3} and \eqref{delta_g}, the channel estimation error $\Delta g_{i^*}$ is given as
\begin{align}\label{delta_g1}
    \Delta g_{i^*} = \mu\left(\frac{\left(z_{i^*}-z_u\right)^{m+1}}{\left\|\hat{\mathbf{u}}-\mathbf{v}_{i^*}+\mathbf{e}_{\mathrm{p}}\right\|^{m+3}} - \frac{\left(z_{i^*}-\hat{z}_{u}\right)^{m+1}}{\left\|\hat{\mathbf{u}}-\mathbf{v}_{i^*}\right\|^{m+3}}\right).
\end{align}
Therefore, $\Delta {g}_{i^*}$ is a function  of ${{\bf{e}}_{\rm{p}}}$, i.e., $\Delta g_{i^*}=f\left(\mathbf{e}_{\mathrm{p}}\right)$ .

Based on the result of positioning and channel estimation, the UE can solve the expected power allocation of each LED to optimize the performance of the integrated VLPC system, and the power allocation result can be fed back to the central controller by the uplink. The designed robust power allocation scheme in this paper will be presented in the following section.

\subsection{Downlink Data Subframe}

During the downlink data subframe, the $i^{*}$-th LED  transmits the data  symbol ${{s_{{\mathrm{c}}}}}$ to the UE, where $ - A \le {s_{\mathrm{c}}} \le A$, $\mathbb{E}\left\{ {{s_{\mathrm{c}}}} \right\} = 0$, and $\mathbb{E}\left\{ {s_{\mathrm{c}}^2} \right\} = \varepsilon $.
Then, the transmitted data signal ${x_{\mathrm{c}}}$ of the ${i^*}$-th LED is given by
\begin{align}\label{x_c}
{x_{\mathrm{c}}} = \sqrt {{P_{\mathrm{c}}}} {s_{\mathrm{c}}} + {I_{{\mathrm{DC}}}},
\end{align}
where ${P_{\mathrm{c}}}$ indicates the allocated   power of the ${i^*}$-th LED. In addition, because the $i^*$-th LED simultaneously transmits the data and positioning signal, the practical transmitted signal can be represented by
\begin{align}
    x_{i^*}=\sqrt{P_{\mathrm{c}}}s_{\mathrm{c}}+\sqrt{P_{\mathrm{p},i}}s_{\mathrm{p},i} + I_{\mathrm{DC}}.
\end{align}
Then, the received data signal ${y_{\mathrm{c}}}$ from the ${i^*}$-th LED is given as
\begin{align}\label{received_data}
{y_{\mathrm{c}}} = \left( {\hat g_{i^*} + \Delta g_{i^*}} \right){x_{\mathrm{c}}} + {n_{\mathrm{c}}},
\end{align}
where ${n_{\mathrm{c}}}\sim \mathcal{N}\left( {0, \sigma _{\mathrm{c}}^2} \right)$ is the received noise, and ${n_{\mathrm{c}}}$ is independent of $\left\{ {{n_{{\mathrm{p}},i}}} \right\}_{i=1}^{M}$.

To ensure that the transmitted  signals of VLC are nonnegative, i.e.,   $x_{{\rm{p}},i}\ge 0$ and   ${x_{i^*}}\ge 0$, the   positioning  signal power ${{P_{{\mathrm{p}},i}}}$ in \eqref{x_p} and the communication signal power ${P_{\mathrm{c}}}$ in \eqref{x_c} are, respectively, constrained by
\begin{subequations}\label{DC_cons}
\begin{align}
&0 \le {P_{{\mathrm{p}},i}} \le \frac{{I_{{\mathrm{DC}}}^2}}
{{{A^2}}},\forall i \in {\tilde{\mathcal{M}}}, i\neq i^*, \\
&0 \le {P_{\mathrm{c}}} + {P_{{\mathrm{p}},i^*}} \le \frac{{I_{{\mathrm{DC}}}^2}}
{{{A^2}}}.
\end{align}
\end{subequations}

For eye safety and practical illumination  requirements, the maximum  optical power of VLC signals should also be limited, i.e.,
${x_{{\rm{p}},i}} \le P_{\rm{o}}^{\max },{x_{i^*}} \le P_{\rm{o}}^{\max }$,
where  $P_{\rm{o}}^{\max }$ denotes the maximum optical power of each LED.  According to \eqref{x_p} and \eqref{x_c},
${{P_{{\mathrm{p}},i}}}$ and  ${P_{\mathrm{c}}}$ are, respectively, limited by
\begin{subequations}\label{opt_max}
\begin{align}
&0 \le {P_{{\rm{p}},i}} \le \frac{{{{\left( {P_{\rm{o}}^{\max } - {I_{{\rm{DC}}}}} \right)}^2}}}{{{A^2}}},\forall i \in {\tilde{\mathcal{M}}}, i\neq i^*,\\
&0 \le {P_{\rm{c}}} + {P_{{\mathrm{p}},i^*}} \le \frac{{{{\left( {P_{\rm{o}}^{\max } - {I_{{\rm{DC}}}}} \right)}^2}}}{{{A^2}}}.
\end{align}
\end{subequations}

Due to the limited power budget of the practical electrical circuit,
the average electrical powers of the signal $x_{\rm{p}}$ and   signal $x_{i^*}$ are also constrained, i.e.,
\begin{subequations}\label{po_power_ele}
\begin{align}
&\mathbb{E}\left\{ x_{\rm{p},i}^2 \right\} = {P_{{\rm{p}},i}}\varepsilon  + I_{{\rm{DC}}}^2 \le {P_{\rm{e}}^{\mathrm{max}}},\forall i \in {\tilde{\mathcal{M}}}, i\neq i^*, \\
&\mathbb{E}\left\{ x_{i^*}^2 \right\} = \left({P_{\mathrm{c}}} + {P_{{\mathrm{p}},i^*}}\right)\varepsilon  + I_{{\rm{DC}}}^2 \le {P_{\rm{e}}^{\mathrm{max}}},
\end{align}
\end{subequations}
where $P_{\rm{e}}^{\mathrm{max}}$ denotes the maximum electrical transmitted power.

By combining constraints \eqref{DC_cons}, \eqref{opt_max} and \eqref{po_power_ele},
${P_{{\mathrm{p}},i}}$ and $ {P_{\mathrm{c}}}$ are constrained by
\begin{subequations}
\begin{align}
&0 \le {P_{{\mathrm{p}},i}} \le \bar{P},\forall i \in {\tilde{\mathcal{M}}}, i\neq i^*,\label{power_constraints_ppi}\\
&0 \le {P_{\mathrm{c}}} + {P_{{\mathrm{p}},i^*}} \le \bar{P},\label{power_constraints_pc}
\end{align}
\end{subequations}
where $\bar{P}_{\mathrm{p}}\triangleq \min \left\{ {\frac{{I_{{\mathrm{DC}}}^2}}
{{{A^2}}},\frac{{P_{\mathrm{e}}^{\mathrm{max}} - I_{{\mathrm{DC}}}^2}}
{\varepsilon },\frac{{{{\left( {P_{\mathrm{o}}^{\max } - {I_{{\mathrm{DC}}}}} \right)}^2}}}
{{{A^2}}}} \right\}$.

Meanwhile, considering the limited load capability in practical circuits, the total power of a VLPC integrated system should be constrained, i.e.,
\begin{align}
    \sum\limits_{i\in\tilde{\mathcal{M}}} P_{\mathrm{p}, i} + P_{\mathrm{c}}\leq P_{\mathrm{total}},\label{total_power_constraint}
\end{align}
where $P_{\mathrm{total}}$ denotes the maximum total power of the VLPC integrated system.

\section{Tradeoff Between Positioning and Communication}\label{sec:tradeoff-between-positioning-and-communication}

In this section, we first present the performance metric for positioning and communication, and then we discuss their tradeoff.

\subsection{CRLB of VLP}

In this paper, we adopt the CRLB as the performance metrics for quantifying the localization accuracy of the UE {\cite{Keskin_2019, BS00,BS01}}.
The CRLB can be achieved by the unbiased estimator since VLC links inherently offer high SNR due to the short transmission distance with the dominant LOS path.

Specifically, we first fix the height of the UE, and focus on the two-dimensional location estimation. Let $\bm{\mathbf{u}} = {\left[ {{x_{\mathrm{u}}},{y_{\mathrm{u}}}} \right]^T}$ denote the arbitrary UE location vector, and the vertical height $z_u$ is a known constant perfectly, i.e., $z_u=\hat{z}_u$.
Based on the received signal \eqref{y_p}, the log-likelihood function of the received signal ${y_{\mathrm{p},i}}$ is given as
\begin{align}\label{ll}
\Lambda \left( \bm{\mathbf{u}} \right) = k - \frac{1}
{{2\sigma _{\mathrm{p}}^2}}\sum\limits_{i\in\tilde{\mathcal{M}}} {\int_0^{{T_{\mathrm{p}}}} {{{\left( {{y_{{\mathrm{p}},i}} - {g_i}\sqrt{P_{\mathrm{p},i}}s_{\mathrm{p},i}} \right)}^2}\,\mathrm{d}t} },
\end{align}
where $k$ is a constant and is independent of $\bm{\mathbf{u}}$.

Moreover, let ${{\mathbf{J}}_{\bm{ \mathbf{u}}}}\left( {{{\mathbf{p}}_{\mathrm{p}}}} \right)$ denote the Fisher Information matrix (FIM) of ${\bm{\mathbf{u}}}$, which is a function of the positioning power ${{\bf{p}}_{\rm{p}}} = {\left[ {{P_{{\rm{p}},1}}, \ldots,{P_{{\rm{p}},M}}} \right]^T}$.
Specifically, the FIM
${{\mathbf{J}}_{\bm{ \mathbf{u}}}}\left( {{{\mathbf{p}}_{\mathrm{p}}}} \right)$ is given as
\begin{align}\label{FIM1}
    {{\mathbf{J}}_{\bm{ \mathbf{u}}}}\left( {{{\mathbf{p}}_{\mathrm{p}}}} \right) = \left[ {\begin{array}{*{20}{c}}
            {\frac{\varepsilon{{T_{\mathrm{p}}}}}
                {{\sigma _{\mathrm{p}}^2}}\sum\limits_{i\in\tilde{\mathcal{M}}} {{P_{{\mathrm{p}},i}}} \frac{{\partial {g_i}}}
                {{\partial {x_{\mathrm{u}}}}}\frac{{\partial {g_i}}}
                {{\partial {x_{\mathrm{u}}}}}} & {\frac{\varepsilon{{T_{\mathrm{p}}}}}
                {{\sigma _{\mathrm{p}}^2}}\sum\limits_{i\in\tilde{\mathcal{M}}}{{P_{{\mathrm{p}},i}}} \frac{{\partial {g_i}}}
                {{\partial {y_{\mathrm{u}}}}}\frac{{\partial {g_i}}}
                {{\partial {x_{\mathrm{u}}}}}}  \\
            {\frac{\varepsilon{{T_{\mathrm{p}}}}}
                {{\sigma _{\mathrm{p}}^2}}\sum\limits_{i\in\tilde{\mathcal{M}}} {{P_{{\mathrm{p}},i}}} \frac{{\partial {g_i}}}
                {{\partial {x_{\mathrm{u}}}}}\frac{{\partial {g_i}}}
                {{\partial {y_{\mathrm{u}}}}}} & {\frac{\varepsilon{{T_{\mathrm{p}}}}}
                {{\sigma _{\mathrm{p}}^2}}\sum\limits_{i\in\tilde{\mathcal{M}}} {{P_{{\mathrm{p}},i}}} \frac{{\partial {g_i}}}
                {{\partial {y_{\mathrm{u}}}}}\frac{{\partial {g_i}}}
                {{\partial {y_{\mathrm{u}}}}}}  \\
    \end{array} } \right],
\end{align}
where
\begin{subequations}
\begin{align}
\frac{{\partial {g_i}}}{{\partial {x_{\rm{u}}}}} =  - \frac{{\left( {m + 3} \right)\mu {{\left( {{z_i} - {z_{\rm{u}}}} \right)}^{m + 1}}\left( {{x_{\rm{u}}} - {x_i}} \right)}}{{{{\left\| {{\bf{u}} - {{\bf{v}}_i}} \right\|}^{m + 5}}}},\\\frac{{\partial {g_i}}}{{\partial {y_{\rm{u}}}}} =  - \frac{{\left( {m + 3} \right)\mu {{\left( {{z_i} - {z_{\rm{u}}}} \right)}^{m + 1}}\left( {{y_{\rm{u}}} - {y_i}} \right)}}{{{{\left\| {{\bf{u}} - {{\bf{v}}_i}} \right\|}^{m + 5}}}}.
\end{align}
\end{subequations}
The derivations of \eqref{FIM1} are given in Appendix A.
Thus, the variance of the zero-mean positioning error ${{\mathbf{e}}_{\mathrm{p}}}$ is lower bounded by  the CRLB {\cite{Keskin_2019}}, i.e.,
\begin{align} \label{FIM}
\mathbb{E}\left\{ {{{\left\| {{{\mathbf{e}}_{\mathrm{p}}}} \right\|}^2}} \right\} \ge \mathrm{Tr}\left( {{\mathbf{J}}_{\bm{ \mathbf{u}}}^{ - 1}\left( {{{\mathbf{p}}_{\mathrm{p}}}} \right)} \right).
\end{align}

\subsection{Achievable Rates of VLC}

Let ${R_{\rm{c}}}$ denote the achievable rate of the UE in the downlink data subframe.
Based on the received signal \eqref{received_data},  ${R_{\rm{c}}}$ is lower bounded by
\begin{subequations}\label{R_c}
    \begin{align}
        {R_{\mathrm{c}}}& = \mathop {\max }\limits_{{f_{\mathrm{c}}}\left( {{s_{\mathrm{c}}}} \right)} I\left( {{x_{\mathrm{c}}};{y_{\mathrm{c}}}} \right)
         = \mathop {\max }\limits_{{f_{\rm{c}}}\left( {{s_{\rm{c}}}} \right)} h\left( {g_{i^*}\sqrt {{P_{\rm{c}}}} {s_{\rm{c}}} + {n_{\rm{c}}}} \right) - h\left( {{n_{\rm{c}}}} \right)\\
        &  \ge \mathop {\max }\limits_{{f_{\mathrm{c}}}\left( {{s_{\mathrm{c}}}\left( t \right)} \right)} W{\log _2}\left( {{2^{2h\left( {\left( {\hat g_{i^*} + \Delta g_{i^*}} \right)\sqrt {{P_{\mathrm{c}}}} {s_{\mathrm{c}}}} \right)}} + {2^{2h\left( {{n_{\rm{c}}}} \right)}}} \right)\nonumber\\
        &\quad\quad- W{\log _2}2\pi e W\sigma _{\mathrm{c}}^2,\label{EPI}  \\
        &   = W{\log _2}\left( {1 + \frac{{{{\left( {\hat g_{i^*} + \Delta g_{i^*}} \right)}^2}{P_{\mathrm{c}}}{e^{1 + 2\left( {\alpha  + \gamma \varepsilon } \right)}}}}{{2\pi W\sigma _{\mathrm{c}}^2}}} \right)= R_{\mathrm{L}}, \label{ABG}
    \end{align}
\end{subequations}
where the inequality \eqref{EPI} holds because of the entropy power inequality (EPI), and \eqref{ABG} holds since ${s_{\rm{c}}}$ follows the ABG distribution that can achieve the maximum differential entropy \cite{Ma}.
Here,  the ABG distribution of  signal ${s_{\rm{c}}}$  is given by
\begin{align}
{f_{\mathrm{c}}}\left( {{s_{\mathrm{c}}}} \right) = \left\{ {\begin{array}{*{20}{c}}
   {{e^{ - 1 - \alpha  - \beta {s_{\mathrm{c}}} - \gamma s_{\mathrm{c}}^2}},\quad - A \le {s_{\mathrm{c}}} \le A;}  \\
   {0,\qquad\qquad\qquad{\mathrm{otherwise}},}  \\
 \end{array} } \right.
\end{align}
where the parameters $\alpha $, $\beta $, $\gamma $ are the solutions of the following equations
\begin{subequations}
    \begin{align}
    &\frac{{\sqrt \pi  {e^{\frac{{{\beta ^2}}}
                    {{4\gamma }}}}\left( {{\rm{erf}}\left( {\frac{{\beta  + 2\gamma A}}
                    {{2\sqrt \gamma  }}} \right) - {\rm{erf}}\left( {\frac{{\beta  - 2\gamma A}}
                    {{2\sqrt \gamma  }}} \right)} \right)}}
    {{2\sqrt \gamma  {e^{1 + \alpha }}}} = 1, \\
    &\beta \left( {{e^{A\left( {\beta  - \gamma A} \right)}} - {e^{ - A\left( {\beta  + \gamma A} \right)}} - {e^{1 + \alpha }}} \right) = 0,\\
    & \frac{{\left( {\beta  - 2\gamma A} \right){e^{ - A\left( {\beta  + \gamma A} \right)}} - \left( {\beta  + 2\gamma A} \right){e^{A\left( {\beta  - \gamma A} \right)}}}}
{{4{\gamma ^2}{e^{1 + \alpha }}}}\nonumber\\
&\qquad\qquad\qquad\qquad\qquad\qquad\qquad + \frac{{{\beta ^2} + 2\gamma }}
{{4{\gamma ^2}}} = \varepsilon,
    \end{align}
\end{subequations}
where ${\rm{erf}}\left( x \right) \triangleq \frac{2}
{{\sqrt \pi  }}\int_0^x {{e^{ - {t^2}}}\,\mathrm{d}t} $ is the error function.


\subsection{Tradeoff Between Positioning and Communication}

The CSI estimation error $\Delta g_{i^*}$   affects the achievable rate ${R_{\rm{c}}}$.
Furthermore,  since $\Delta {g}_{i^*} = f\left( {{{\mathbf{e}}_{\mathrm{p}}}} \right)$  and  ${{\bf{e}}_{\rm{p}}}$  depend on the positioning  signal power $\left\{ {{P_{{\rm{p}},i}}} \right\}$,
the achievable rate ${R_{\rm{c}}}$ not only depends on the communication signal power ${P_{\mathrm{c}}}$, but also
depends on the positioning  signal power $\left\{ {{P_{{\rm{p}},i}}} \right\}$.
Therefore, there exists an optimal tradeoff between communication power ${P_{\mathrm{c}}}$ and positioning power ${{P_{{\mathrm{p}},i}}}$ for the  integrated VLPC system design.

\section{Robust Integrated VLPC Design}\label{sec:robust-integrated-vlpc-design}

For any given distribution of the positioning errors, it is hard to design an integrated VLPC scheme for the UE, which always meets the positioning or throughput requirements due to the unbounded errors.
However, it is reasonable to make a robust design within the tolerance of uncertainty in practice.
In this section, we investigate a chance-constrained robust integrated VLPC design for the  two types  of  positioning error  distributions: Gaussian distribution and arbitrary distribution.
Specifically, by exploiting  the relationship between positioning errors and CSI errors, we focus on designing a VLPC power allocation scheme  to minimize the CRLB of VLP  under the QoS chance constraint and power constraints.

\subsection{VLPC Design With Prefect CSI}
Under the minimum rate requirement, the optical and electrical power constraints, the optimal VLPC power allocation  to minimize the CRLB of VLP with perfect CSI can be formulated as follows:
\begin{subequations}\label{VLPC_problem_perfect }
    \begin{align}
        \min_{\mathbf{p}_{\mathrm{p}}, P_{\mathrm{c}}} & \mathrm{Tr}\left(\mathbf{J}_{\mathbf{u}}^{-1}\left(\mathbf{p}_{\mathrm{p}}\right)\right)\\
        \mathrm{s.t.}&\eqref{power_constraints_ppi}, \eqref{power_constraints_pc}, \eqref{total_power_constraint}\nonumber\\
        &R_\mathrm{c}\geq\bar{r},\label{min_rate_constraint}
    \end{align}
\end{subequations}
where $\bar{r}$ denotes the minimum rate requirement. After replacing the constraint \eqref{min_rate_constraint} with a more stringent constraint $R_{\mathrm{L}}\geq  \bar{r}$,
problem \eqref{VLPC_problem_perfect } can be rewritten as a convex programming problem, which can be solved by  standard optimization solvers such as CVX \cite{cvx}. In addition, \eqref{VLPC_problem_perfect } also corresponds to the nonrobust design ignoring the coupling between the positioning error $\mathbf{e}_{\mathrm{p}}$ and the estimated CSI error $\Delta g$. The optimal communication power and positioning power are denoted by $P_{\text{c}}^*$ and $\mathbf{p}_{\text{p}}^*$, respectively.

\subsection{Robust VLPC Design with Gaussian Distributed ${{{\bf{e}}_{\rm{p}}}}$}\label{sec:robust-VLPC-design-with-gaussian-distributed-bfermp}

Considering the relationship between $\Delta  g$ and $\mathbf{e}_{\mathrm{p}}$, we propose the outage chance  constraint to handle the minimum rate requirement \eqref{min_rate_constraint}.  When $\mathbf{e}_{\mathrm{p}}$ follows the Gaussian distribution, the corresponding  chance-constrained VLPC programming problem can be stated as
\begin{subequations}\label{VLPC_problem}
    \begin{align}
        \min_{\mathbf{p}_{\mathrm{p}},P_{\mathrm{c}}} &\mathrm{Tr}\left(\mathbf{J}_{\mathbf{u}}^{-1}\left(\mathbf{p}_{\mathrm{p}}\right)\right)\\
        \mathrm{s.t.}&\eqref{power_constraints_ppi}, \eqref{power_constraints_pc}, \eqref{total_power_constraint}\nonumber\\
        &\mathrm{Pr}\left\{R_\mathrm{c}\leq \bar{r}\right\}\leq P_{\mathrm{out}},\label{outage_probability}\\
        &\mathbf{e}_{\mathbf{p}}\sim\mathcal{N}\left(\mathbf{0},\mathbf{J}_{\mathbf{u}}^{-1}\left(\mathbf{p}_{\mathrm{p}}\right)\right),\label{VLPC_problem_e_p}
    \end{align}
\end{subequations}
where $P_{\mathrm{out}}$ denotes the maximum tolerable outage probability.

The robust integrated VLPC design problem \eqref{VLPC_problem} is nonconvex and computationally intractable. The main challenge lies in the chance  constraint \eqref{outage_probability}, which does not admit closed-form expressions.
In order to handle the chance-constrained problem \eqref{VLPC_problem}, we first reformulate  the chance  constraint  \eqref{outage_probability}.
Based on the lower bound of the achievable rate \eqref{ABG}, the probability  constraint \eqref{outage_probability}
can be conservatively transformed into the following constraint
\begin{align}
    \mathrm{Pr}\left\{R_\mathrm{L}\leq \bar{r}\right\}\leq P_{\mathrm{out}}.\label{outage_probability_2}
\end{align}
Specifically, the constraint $R_{\mathrm{L}}\geq  \bar{r}$ can be equivalently reformulated as
\begin{align}\label{Gauss_1}
{\left\| {{\bf{u}} - {{\bf{v}}_{{i^*}}}} \right\|^2} \ge P_c^{\frac{1}{{{m + 3}}}}\delta,
\end{align}
where $\delta  \triangleq {\left( {\frac{{{e^{1 + 2\left( {\alpha  + \gamma \varepsilon } \right)}}{\mu ^2}{{\left( {{z_{{i^*}}} - {z_{\mathrm{u}}}} \right)}^{2\left( {m + 1} \right)}}}}
{{2\pi W \sigma _{\mathrm{c}}^2\left( {{2^{{\frac{\bar R}{W}}}} - 1} \right)}}} \right)^{\frac{1}
{{ {m + 3} }}}}$.

Then, by substituting ${\mathbf{u}} = {\mathbf{\hat u}} + {{\mathbf{e}}_{\mathrm{p}}}$ into \eqref{Gauss_1}, we have
 \begin{align}
{\left\| {{{\bf{e}}_{\rm{p}}}} \right\|^2} + 2{\bf{e}}_{\rm{p}}^T\left( {{\bf{\hat u}} - {{\bf{v}}_{{i^*}}}} \right) \ge \delta {P_{\rm{c}}}^{\frac{1}{{m + 3}}}  - {{{\left\| {{\bf{\hat u}} - {{\bf{v}}_{{i^*}}}} \right\|}^2}}.
\end{align}

Moreover, the positioning error ${{\bf{e}}_{\rm{p}}}$ can be rewritten as $\mathbf{e}_\mathrm{p} = \mathbf{J}_{\mathbf{u}}^{-\frac{1}{2}}\left(\mathbf{p}_{\mathrm{p}}\right)\mathbf{\hat{e}}_{\mathrm{p}}$,
where ${{\mathbf{\hat e}}_{\mathrm{p}}}\sim \mathcal{N}\left( {{\mathbf{0}},{\mathbf{I}}} \right)$.
Then, the chance   constraint  \eqref{outage_probability} can be reformulated as
\begin{align}  \label{Gauss_3}
{\rm{Pr}}\left\{ {{\bf{\hat e}}_{\rm{p}}^T{\bf{B}}{{{\bf{\hat e}}}_{\rm{p}}} + 2{\bf{\hat e}}_{\rm{p}}^T{\bf{b}} \ge {\delta _b}} \right\} \le {P_{{\rm{out}}}},
\end{align}
where $\mathbf{B} \triangleq\mathbf{J}_{\mathbf{u}}^{- 1}\left( \mathbf{p}_{\mathrm{p}}\right)$,
 $\mathbf{b} \triangleq \mathbf{J}_{\mathbf{u}}^{-\frac{1}{2}} \left(\mathbf{p}_{\mathrm{p}}\right)\left(\mathbf{\hat{u}} - \mathbf{v}_{i^*} \right)$,
 and ${\delta _b} = \delta P_{\mathrm{c}}^{\frac{1}{m+3}}  - {{{\left\| {{\bf{\hat u}} - {{\bf{v}}_{{i^*}}}} \right\|}^2}}$.

Furthermore, to reformulate the   chance  constraint \eqref{Gauss_3} into a deterministic form, we invoke the    Bernstein-type inequality  in   Lemma $1$.

\textbf{Lemma $1$}{\cite{Wang_2014} \cite{Ma_2017}}: Let $L  = \mathbf{x}^T\mathbf{B}\mathbf{x} + 2\mathbf{x}^T\mathbf{b}$, where ${\mathbf{B}} \in {{\mathbf{R}}^{N \times N}}$ is a real symmetric matrix, ${\mathbf{b}} \in {\mathbb{R}^{N \times 1}}$ and  ${\bf{x}} \sim {\mathcal{N}}\left( {0,{\bf{I}}} \right)$. Then, for any $\eta  \ge 0$, we have
\begin{align}
&{\rm{Pr}}\left\{ {L  \ge {\mathrm{Tr}}\left( {\mathbf{B}} \right)   +   \sqrt {{\mathrm{2}}\eta } \sqrt {{{\left\|\mathbf{B} \right\|}_{\mathrm{F}}^{\mathrm{2}}}   +  2 {{\left\| {\mathbf{b}} \right\|}^{\mathrm{2}}}} +  \eta {\lambda^+ }\left( {\mathbf{B}} \right)} \right\}\nonumber\\
&\qquad\qquad\qquad\qquad\qquad\qquad\qquad\qquad\le \exp \left( { - \eta } \right),
\end{align}
where ${\lambda^+ }\left( {\mathbf{B}} \right) = \max \left\{ {{\lambda _{\max }}\left( {\mathbf{B}} \right),0} \right\}$ and ${\lambda _{\max }}\left( {\mathbf{B}} \right)$ is the maximum eigenvalue of matrix ${\mathbf{B}}$.

According to the Bernstein-type inequality, the  chance  constraint \eqref{Gauss_3} can be conservatively   transformed into the following constraint
\begin{align} \label{Gauss_4}
{\rm{Tr}}\left( {\bf{B}} \right) + \sqrt {{\rm{2}}\eta } \sqrt {{{\left\|\bf{B}\right\|}_{\mathrm{F}}^2} + 2{{\left\| {\bf{b}} \right\|}^2}}  + \eta {\lambda^ + }\left( {\bf{B}} \right) \le {\delta _b},
\end{align}
where $\eta  =  - \ln \left( {{P_{{\rm{out}}}}} \right)$.
Furthermore,   constraint \eqref{Gauss_4} can be   equivalently reformulated  as
\begin{subequations}
    \begin{align}
        &\mathrm{Tr}\left(\mathbf{B}\right)+\sqrt{2\eta}\omega+\eta \varrho - \delta_b\leq 0,\label{gaus_outage_1}\\
        &\begin{Vmatrix}
            \mathrm{vec}\left(\mathbf{B}\right)\\
            \sqrt{2}\mathbf{b}
        \end{Vmatrix}\leq \omega,\label{gaus_outage_2}\\
        &\varrho\mathbf{I} - \mathbf{B}\succeq \mathbf{0}, \varrho\geq 0,\label{gaus_outage_3}.
    \end{align}
\end{subequations}
where $\omega  $ and $\rho$ are slack variables.

Note that constraint \eqref{gaus_outage_1} is convex, but  constraints \eqref{gaus_outage_2} and \eqref{gaus_outage_3} are nonconvex because the matrix inverse is non-convexity-preserving.
To tackle this issue, constraints \eqref{gaus_outage_2} and \eqref{gaus_outage_3} can be transformed to convex forms by exploiting the eigenvalue and singular value properties.

Specifically, the left-hand side of constraint \eqref{gaus_outage_2}  can be   approximated  as
\begin{align}
    \left\|\begin{array}{c}\mathrm{vec}\left(\mathbf{B}\right)\\ \sqrt{2}\mathbf{b}\end{array}\right\|
    &\leq\left\|\mathbf{B}\right\|_{\mathrm{F}}+\sqrt{2}\left\|\mathbf{b}\right\|\nonumber\\
    &\leq \sqrt{\mathrm{rank\left(\mathbf{B}\right)}}\sigma_{\mathrm{max}}\left(\mathbf{B}\right)+\sqrt{2}\sigma_{\mathrm{max}}\left(\mathbf{b}\right),
\end{align}
where $\sigma_{\mathrm{max}}\left(\cdot\right)$ denotes the maximum singular value, and the first inequality  is due to $\sqrt{a^2+b^2}\leq a +b, a\geq0,b\geq0$, and the second inequality is because the norm is equal to the sum of all singular values. Because the FIM is invertible and semidefinite, we have
\begin{subequations}
    \begin{align}
        \mathrm{rank\left(\mathbf{B}\right)}&=2,\\
        \sigma_{\mathrm{max}}\left(\mathbf{B}\right)&=\lambda_{\mathrm{min}}^{-1}\left(\mathbf{J}_{\mathbf{u}}\left(\mathbf{p}_{\mathrm{p}}\right)\right),\\
        \sigma_{\mathrm{max}}\left(\mathbf{b}\right)&\leq \sigma_{\mathrm{max}}\left(\mathbf{B}^{\frac{1}{2}}\right)\sigma_{\mathrm{max}}\left(\hat{\mathbf{u}}-\mathbf{v}_{i^*}\right)\nonumber\\
        &=\frac{\sigma_{\mathrm{max}}\left(\hat{\mathbf{u}}-\mathbf{v}_{i^*}\right)}{\sqrt{\lambda_{\mathrm{min}}\left(\mathbf{J}_{\mathbf{u}}\left(\mathbf{p}_{\mathrm{p}}\right)\right)}},
    \end{align}
\end{subequations}
where $\lambda_{\min}\left(\cdot\right)$ denotes the minimum eigenvalue of the matrix.
Thus, an upper bound on the left-hand side of constraint \eqref{gaus_outage_2} can be derived, and constraint \eqref{gaus_outage_2} can be rewritten with a convex form as follows
\begin{align}
    \left\|\begin{array}{c}\mathrm{vec}\left(\mathbf{B}\right)\\ \sqrt{2}\mathbf{b}\end{array}\right\|
    &\leq \frac{\sqrt{2}}{\lambda_{\mathrm{min}}\left(\mathbf{J}_\mathbf{u}\left(\mathbf{p}_{\mathrm{p}}\right)\right)}+\frac{\sqrt{2}\sigma\left(\hat{\mathbf{u}}-\mathbf{v}_{i^*}\right)}{\sqrt{\lambda_{\mathrm{min}}\left(\mathbf{J}_\mathbf{u}\left(\mathbf{p}_{\mathrm{p}}\right)\right)}} \leq \omega.\label{gaus_outage_2_appr_2}
\end{align}

Meanwhile, because the positive semidefinite matrix means that its minimum eigenvalue is nonnegative, constraint \eqref{gaus_outage_3} can be transformed to a concave form as follows
\begin{align}
    \lambda_{\mathrm{min}}\left(\rho\mathbf{I}-\mathbf{J}_\mathbf{u}^{-1}\left(\mathbf{p}_{\mathrm{p}}\right)\right)=\rho-\frac{1}{\lambda_{\mathrm{min}}\left(\mathbf{J}_\mathbf{u}\left(\mathbf{p}_{\mathrm{p}}\right)\right)}\geq 0.\label{gaus_outage_3_eig}
\end{align}

Thus, problem \eqref{VLPC_problem} can be  reformulated as
    \begin{align} \label{gauss_last}
    \min_{\mathbf{p}_{\mathrm{p}},P_{\mathrm{c}}} &~\mathrm{Tr}\left(\mathbf{J}_{\mathbf{u}}^{-1}\left(\mathbf{p}_{\mathrm{p}}\right)\right)\\
    \rm{s.t.}&~\eqref{power_constraints_ppi}, \eqref{power_constraints_pc}, \eqref{total_power_constraint}, \eqref{gaus_outage_1}, \eqref{gaus_outage_2_appr_2},
    \eqref{gaus_outage_3_eig}.\nonumber
    \end{align}
Problem \eqref{gauss_last} is a convex semidefinite program (SDP), which   can be solved by standard optimization solvers such as CVX\cite{cvx}.

\subsection{Robust VLPC Design with Arbitrary Distributed ${{{\bf{e}}_{\rm{p}}}}$}\label{sec:robust-VLPC-design-with-arbitrary-distributed-bfermp}

In this subsection, we investigate a more practical robust VLPC design scenario, where the central controller has no prior knowledge of the distribution of the position error   ${{\bf{e}}_{\rm{p}}}$
except for its first and second-order moments, i.e., only the mean   and variance of  ${{\bf{e}}_{\rm{p}}}$ are known.
Specifically, although  the distribution of ${{\bf{e}}_{\rm{p}}}$ is arbitrary,
the positioning error variance can achieve the CRLB, i.e., $\mathbb{E}\left\{\mathbf{e}_{\mathrm{p}}\mathbf{e}_{\mathrm{p}}^{T}\right\}=\mathbf{J}_\mathbf{u}^{-1}\left(\mathbf{p}_{\mathrm{p}}\right)$, and the mean of ${{{\bf{e}}_{\rm{p}}}}$ is zero, i.e., $\mathbb{E}\left\{ {{{\bf{e}}_{\rm{p}}}} \right\}  =  0$.

With the arbitrary distributed ${{\bf{e}}_{\rm{p}}}$,
we aim to minimize the  CRLB of VLP by optimizing the power allocation subject to the VLC chance constraint and power constraints.
Mathematically, the robust VLPC problem can be formulated as follows:
\begin{subequations}\label{VLPC_problem_arbitrary}
    \begin{align}
        \min_{\mathbf{p}_{\mathrm{p}},P_{\mathrm{c}}} &\mathrm{Tr}\left(\mathbf{J}_{\mathbf{u}}^{-1}\left(\mathbf{p}_{\mathrm{p}}\right)\right)\\
        \mathrm{s.t.}&\mathbb{E}\left\{\mathbf{e}_{\mathrm{p}}\mathbf{e}_{\mathrm{p}}^{T}\right\}=\mathbf{J}_{\mathbf{u}}^{-1}\left(\mathbf{p}_{\mathrm{p}}\right),\mathbb{E}\left\{ {{{\bf{e}}_{\rm{p}}}} \right\}{\rm{ = }}0,\label{VLPC_problem_arbitrary_e_p_1}\\
        &\eqref{outage_probability_2},\eqref{total_power_constraint}, \eqref{power_constraints_ppi}, \eqref{power_constraints_pc}.\nonumber
    \end{align}
\end{subequations}

Problem \eqref{VLPC_problem_arbitrary} appears to be more challenging than
problem \eqref{VLPC_problem} since less information about the distribution of ${{\bf{e}}_{\rm{p}}}$ is known.
To reformulate the intractable chance constraints to computationally tractable constraints, \eqref{outage_probability} can be equivalently expressed as
\begin{align}\label{R_c_quad_inv}
    &\operatorname{Pr}\left\{\hat{\mathbf{e}}_{\text{p}}^T \mathbf{B}\hat{\mathbf{e}}_{\text{p}} + 2\mathbf{b}^T \hat{\mathbf{e}}_{\text{p}} -\delta_b \leq 0\right\}
    \geq 1 - P_{\text{out}}.
\end{align}
Furthermore, the chance constraint \eqref{R_c_quad_inv} can be transformed into a distributionally robust chance constraint, which is given by
\begin{align}\label{distributionally_robust_chance_constraint}
    &\inf_{\mathbb{P}\in\mathcal{P}} \operatorname{Pr}\left\{\hat{\mathbf{e}}_{\text{p}}^T \mathbf{B}\hat{\mathbf{e}}_{\text{p}} + 2\mathbf{b}^T \hat{\mathbf{e}}_{\text{p}} -\delta_b \leq 0 \right\} \geq 1 - P_{\text{out}},
\end{align}
where $\inf_{\mathbb{P}\in\mathcal{P}} \operatorname{Pr}\left\{ \cdot \right\}$ denotes the probability lower bound under the probability distribution $\mathbb{P}$ and $\mathcal{P}$ is called the ambiguity set, which includes all the possible position error distributions.

\textbf{Lemma $2$}\cite{ZYMLER_CVaR,ZHANG_IOTJ_CVaR}: Consider a continuous loss function
$L:\mathbb{R}^{k}\to\mathbb{R}$ that is concave or quadratic in
$\bm{\xi}$. The distributionally robust chance constraint is equivalent to the worst-case constraint, which is given by
\begin{align}
\inf_{\mathbb{P}\in\mathcal{P}}\operatorname{Pr}_{\mathbb{P}}&\left\{L\left( \bm{\xi}\right)\leq0\right\}\geq1-\epsilon\Leftrightarrow\sup_{\mathbb{P}\in\mathcal{P}}\operatorname{\mathbb{P}-CVaR}_{\epsilon}\left\{L\left( \bm{\xi}\right)\right\}\leq0,
\end{align}
where
$\operatorname{\mathbb{P}-CVaR}_{\epsilon}\left\{L\left( \bm{\xi}\right)\right\} $ is denoted as the CVaR of  $L\left( \bm{\xi}\right) $ at the threshold  $\epsilon $ with respect to  $\mathbb{P} $, defined as
\begin{align}
\operatorname{\mathbb{P}-CVaR}_{\epsilon}\left\{L\left( \bm{\xi}\right)\right\}=\inf_{\beta\in\mathbb{R}}\left\{\beta+\frac{1}{\epsilon}\mathbb{E}_{\mathbb{P}}\left[\left(L\left( \bm{\xi}\right)-\beta\right)^+\right]\right\}.
\end{align}
Moreover,  $\mathbb{R} $ is the set of real numbers and  $\left(z\right)^+=\max\left\{0,z\right\} $, and  $\beta\in\mathbb{R} $
is an auxiliary variable introduced by the CVaR.

\textbf{Lemma $3$} \cite{ZYMLER_CVaR,ZHANG_IOTJ_CVaR}: Let  $L\left( \bm{\xi}\right)=  \bm{\xi}^T \mathbf{Q} \bm{\xi}+\mathbf{q}^T  \bm{\xi}+q^0 $ be a quadratic function of  $ \bm{\xi} $,  $\forall \bm{\xi}\in\mathbb{R}^{n} $. The worst-case CVaR can be computed as
\begin{subequations}
\begin{align}
\sup_{\mathbb{P}\in\mathcal{P}}&\, \operatorname{\mathbb{P}-CVaR}_{\epsilon}\left\{ L\left( \bm{\xi}\right) \right\} = \min_{\beta,\mathbf{M}} \beta+\frac{1}{\epsilon}\mathrm{Tr}\left(\mathbf{\Omega}\mathbf{M}\right)\\
\mathrm{s.t.}&\, \mathbf{M}\in\mathbb{S}^{n+1},\mathbf{M}\succeq\mathbf{0},\\
&\, \mathbf{M}-
\begin{bmatrix}
\mathbf{Q} & \frac{1}{2}\mathbf{q}\\
\frac{1}{2}\mathbf{q}^T & q^0-\beta
\end{bmatrix}
\succeq \mathbf{0},
\end{align}
\end{subequations}
where  $\mathbf{M} $ and  $\beta\in\mathbb{R}$ are the auxiliary variables,
and  $\mathbf{\Omega} $ is a matrix defined as
\begin{align}
\mathbf{\Omega}\triangleq
\begin{bmatrix}
\mathbf{\Sigma}+ \bm{\mu} \bm{\mu}^T &  \bm{\mu}\\
\bm{\mu}^T & 1
\end{bmatrix},
\end{align}
where  $\bm{\mu}\in\mathbb{R}^n $ and $\mathbf{\Sigma}\in\mathbb{S}^n $ are the mean and covariance
matrix of random vector  $ \bm{\xi} $, respectively.

The arbitrary distributed positioning error $\mathbf{e}_{\mathrm{p}}$ makes the chance constraint lower bound intractable. However, a CVaR-based method can overcome this challenge effectively, which is known as a good convex approximation of the worst-case chance constraint\cite{ZYMLER_CVaR,ZHANG_IOTJ_CVaR}. As is described by Lemmas 2 and 3, for the continuous quadratic function $L\left(\hat{\mathbf{e}}_{\mathrm{p}}\right)\triangleq \hat{\mathbf{e}}_{\text{p}}^T \mathbf{B}\hat{\mathbf{e}}_{\text{p}} + 2\mathbf{b}^T \hat{\mathbf{e}}_{\text{p}} -\delta_b$, the distributionally robust chance constraint \eqref{distributionally_robust_chance_constraint} can be made equivalent to the worst-case CVaR constraint as follows:
\begin{subequations}\label{CVaR_constraint}
\begin{align}
            &\beta+\frac{1}{P_{\mathrm{out}}}\mathrm{Tr}\left(\mathbf{\Omega}\mathbf{M}\right)\leq 0;\label{CVaR_constraint_a}\\
            &\mathbf{M}\in\mathbb{S}^{3},\mathbf{M}\succeq0;\label{CVaR_constraint_b}\\
&\mathbf{M}-
            \begin{bmatrix}
                \mathbf{B} & \mathbf{b}\\
                \mathbf{b}^T & -\delta_b - \beta
            \end{bmatrix}\succeq \mathbf{0},\label{CVaR_constraint_e}
\end{align}
\end{subequations}
where $\mathbf{M}$ and $\beta\in\mathbb{R}$ are two auxiliary variables, and $\mathbf{\Omega}=
\begin{bmatrix}
\mathbf{I} & \mathbf{0} \\
\mathbf{0}^{T} & 1
\end{bmatrix}$.

The successive convex approximation (SCA) based on the first-order Taylor expansion can be exploited to process the nonconvex constraint  \eqref{CVaR_constraint_e}.
The first-order Taylor expansion of the terms $\mathbf{J}_{\mathbf{u}}^{-1}\left(\mathbf{p}_{\mathrm{p}}\right)$,  $\mathbf{J}_{\mathbf{u}}^{-\frac{1}{2}}\left(\mathbf{p}_{\mathrm{p}}\right)$ and $P_{\mathrm{c}}^{\frac{1}{m+3}}$ is substituted into \eqref{CVaR_constraint_e} to find an affine approximation, which is given by
\begin{align}
    \mathbf{M}-
    \begin{bmatrix}
        \tilde{\mathbf{B}} & \tilde{\mathbf{b}}\\
        \tilde{\mathbf{b}}^T & -\tilde{\delta}_b - \beta
    \end{bmatrix}\succeq \mathbf{0},\label{CVaR_constraint_e_convexity}
\end{align}
where $\tilde{\mathbf{B}}$, $\tilde{\mathbf{b}}$ and $\tilde{\delta}_b$ denote the approximations of ${\mathbf{B}}$, ${\mathbf{b}}$ and ${\delta}_b$ through
\begin{subequations}
\begin{align}
    &\mathbf{J}_{\mathbf{u}}^{-1}\left(\mathbf{p}_{\mathrm{p}}\right)\approx\mathbf{J}_{\mathbf{u}}^{-1}\left(\mathbf{p}_{\mathrm{p},0}\right)\nonumber\\
    &-\mathbf{J}_{\mathbf{u}}^{-1}\left(\mathbf{p}_{\mathrm{p},0}\right)\left(\mathbf{J}_{\mathbf{u}}\left(\mathbf{p}_{\mathrm{p}}\right)-\mathbf{J}_{\mathbf{u}}\left(\mathbf{p}_{\mathrm{p},0}\right)\right)\mathbf{J}_{\mathbf{u}}^{-1}\left(\mathbf{p}_{\mathrm{p},0}\right),\label{inv_FIM_taylor_appro}\\
&\mathbf{J}_{\mathbf{u}}^{-\frac{1}{2}}\left(\mathbf{p}_{\mathrm{p}}\right)\approx \mathbf{J}_{\mathbf{u}}^{-\frac{1}{2}}\left(\mathbf{p}_{\mathrm{p},0}\right)\nonumber\\
    &-\frac{1}{2}\mathbf{J}_{\mathbf{u}}^{-\frac{3}{4}}\left(\mathbf{p}_{\mathrm{p},0}\right)\left(\mathbf{J}_{\mathbf{u}}\left(\mathbf{p}_{\mathrm{p}}\right)-\mathbf{J}_{\mathbf{u}}\left(\mathbf{p}_{\mathrm{p},0}\right)\right)\mathbf{J}_{\mathbf{u}}^{-\frac{3}{4}}\left(\mathbf{p}_{\mathrm{p},0}\right),\label{sqrtm_inv_FIM_taylor_appro}\\
&P_{\mathrm{c}}^{\frac{1}{m+3}}\approx P_{\mathrm{c},0}^{\frac{1}{m+3}} +\frac{1}{m+3}P_{\mathrm{c},0}^{-\frac{m+2}{m+3}}\left(P_{\mathrm{c}}-P_{\mathrm{c},0}\right).\label{P_c_pow_taylor_appro}
\end{align}
\end{subequations}

Thus, the chance-constrained problem \eqref{VLPC_problem_arbitrary} can be reformulated as follows
\begin{align}\label{arb_opt_problem}
    \min_{\mathbf{p}_{\mathrm{p}},P_{\mathrm{c}}, \mathbf{M}, \beta}&\mathrm{Tr}\left(\mathbf{J}_{\mathbf{u}}^{-1}\left(\mathbf{p}_{\mathrm{p}}\right)\right)\\
    \mathrm{s.t.}&~\eqref{power_constraints_ppi}, \eqref{power_constraints_pc}, \eqref{total_power_constraint},\eqref{CVaR_constraint_a}, \eqref{CVaR_constraint_b},\eqref{CVaR_constraint_e_convexity}.\nonumber
\end{align}
For any known $\mathbf{p}_{\mathrm{p},0}$ and $P_{\mathrm{c},0}$, the joint optimization problem \eqref{arb_opt_problem} becomes the convex SDP, which can be solved by standard convex programming solvers such as CVX\cite{cvx}. Toward this end, we transform problem \eqref{arb_opt_problem} into a series of convex subproblems, which can be solved efficiently through iterations. At the $t$-th iteration, the corresponding convex subproblem is given as
\begin{subequations}\label{arb_opt_problem_sequential_problem}
\begin{align}\label{arb_opt_problem_sequential}
    \min_{\mathbf{p}_{\mathrm{p}},P_{\mathrm{c}}, \mathbf{M}, \beta}&\mathrm{Tr}\left(\mathbf{J}_{\mathbf{u}}^{-1}\left(\mathbf{p}_{\mathrm{p}}\right)\right)\\
    \mathrm{s.t.} & \mathrm{Tr}\left(\mathbf{J}_{\mathbf{u}}^{-1}\left(\mathbf{p}_{\mathrm{p}}\right)\right)\leq c^{\left(t-1\right)}\\
    &~\eqref{power_constraints_ppi}, \eqref{power_constraints_pc}, \eqref{total_power_constraint},\eqref{CVaR_constraint_a}, \eqref{CVaR_constraint_b}, \eqref{CVaR_constraint_e_convexity}.\nonumber
\end{align}
\end{subequations}
Then, the iterations repeat until the termination condition is satisfied, and the optimal solutions $\mathbf{p}_{\mathrm{p},0}$ and $P_{\mathrm{c},0}$ are output. The details of the robust integrated VLPC method for the arbitrary distributed $\mathbf{e}_{\mathrm{p}}$ are summarized as Algorithm~\ref{alg:Framwork_unknown}.
The proposed SCA algorithm can converge to a stationary point of the original problem after using the CVaR-based method\cite{zhouSolvingHighOrderPortfolios2021a}. At each iteration, the subproblem (50) can be efficiently solved with a worst-case complexity $\mathcal{O}\left(\left(M+11\right)^{4.5}\log\left(1/\delta\right)\right)$, where $\delta>0$ is the accuracy of the interior-point method\cite{cvx,luoSemidefiniteRelaxationQuadratic2010}.

\begin{algorithm}
    \caption{Robust integrated VLPC  for the arbitrary distributed $\mathbf{e}_{\mathbf{p}}$.}\label{alg:Framwork_unknown}
    \begin{algorithmic}[1]
        \Require Given ${\epsilon} \ge 0$, $t = 1$, $c^{\left(0\right)} = \infty$, choose proper $\mathbf{p}_{\mathrm{p},0}$, $\mathbf{P}_{\mathrm{c},0}$, and set $ \bar{r}$ and, $P_{\mathrm{out}}$;
        \Repeat
        \State{Solve   problem \eqref{arb_opt_problem_sequential} to obtain  $\mathbf{p}_{\mathrm{p}},P_{\mathrm{c}}$, and calculate $\mathrm{CRLB}$;}
        \State Set $\mathbf{p}_{\mathrm{p},0} = \mathbf{p}_{\mathrm{p}}$, $\mathbf{P}_{\mathrm{c},0} = \mathbf{P}_{\mathrm{c}}$, $c^{\left(t\right)} =  \mathrm{CRLB}$;
        \State $t = t+1$.
        \Until{$\frac{\left|c^{\left(t\right)}-c^{\left(t-1\right)}\right|}{c^{\left(t\right)}}\leq \epsilon$.}
        \Ensure Output the optimal solutions ${{\mathbf{p}}_{\mathrm{p}}}$ and ${P_{\mathrm{c}}}$.
    \end{algorithmic}
\end{algorithm}

\section{numerical results}\label{sec:numerical-results}

\begin{figure}[h]
    \centering
    \includegraphics[width=0.25\textwidth]{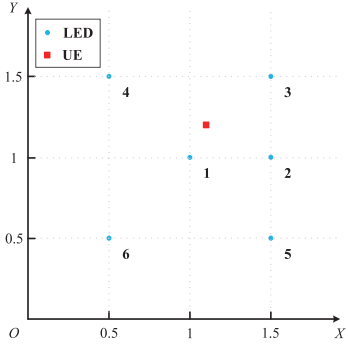}
    \caption{~The location of UE and LEDs.}
    \label{fig_location}
\end{figure}

This section presents numerical results to show the proposed robust power allocation schemes for the integrated VLPC system. We consider an indoor VLPC system installed with multiple LEDs {on the ceiling}, where {the room height is $2.5~\mathrm{m}$, and} a corner of a square room denotes the origin $(0, 0, 0)$ of a three-dimensional Cartesian coordinate system $(X, Y, Z)$.  The receiver's location is $(1.1, 1.2, 1.5)\,\mathrm{m}$, and four numbers  of transmitters are considered, namely, 3 LEDs, 4 LEDs, 5 LEDs, and 6 LEDs, where their locations are shown in Fig. \ref{fig_location}. According to the channel model \eqref{channel_gain}, it can be verified that all LEDs are within the UE's FoV. The signal $s_{\mathrm{c}}$ is assumed to be drawn from a uniform distribution  $\mathcal{U}\left(-0.1, 0.1\right) $, i.e., $A = 0.1$, $\alpha = \ln 2\sqrt{0.1} - 1, \beta = \gamma = 0$. Without loss of generality, we assume that the total power is $P_{\mathrm{total}}=3\min\left\{\bar{P}_{\mathrm{p}},\bar{P}_{\mathrm{c}}\right\}$. According to  \cite{Yang_TWC_2020,Yang_IOT_2020,Keskin_2019}, the other simulation parameters are listed in TABLE \ref{baisc_par}.

\begin{table}[h]
    \centering
    \caption{Simulation Parameters.}\label{baisc_par}
    \begin{tabular}{|l|l|}
        \hline
        Definition  & Value   \\
        \hline
        Lambertian index, $m$ & $1$   \\
        \hline
        Angle of FoV, $\psi _{\mathrm{FoV}}$ & ${120^ \circ }$   \\
        \hline
        Half power angle, ${\Phi _{1/2}}$&${60^ \circ }$ \\
        \hline
        PD effective area, $A_{\mathrm{R}}$ & $1~\mathrm{cm}^2$  \\
        \hline
        Conversion coefficient, $\eta_{\mathrm{\ell}}$, $\eta_{\mathrm{c}}$ & $1$  \\
        \hline
        Bandwidth, $W$ & $20~\mathrm{MHz}$\\
        \hline
        DC bias, $I_{\mathrm{DC}}$, & $1~\mathrm{A}$ \\
        \hline
        Noise power, ${\sigma_{\mathrm{p}} ^2}$, ${\sigma_{\mathrm{c}} ^2}$ & $10^{-22}~\rm{A^2/Hz}$\\
        \hline
        Maximum optical power, $P_{\mathrm{o}}^{\mathrm{max}}$ & $5$ W\\
        \hline
        Maximum electrical power, $P_{\mathrm{e}}^{\mathrm{max}}$ & $5$ W\\
        \hline
        Positioning subframe length, $T_{\mathrm{p}}$ & $0.1$ $\mu$s\\
        \hline
    \end{tabular}
\end{table}

\subsection{Cumulative Distribution Functions of Communication Rates}

\begin{figure}[htbp]
    \centering
    \begin{minipage}[b]{0.35\textwidth}
        \centering
        \includegraphics[width =\textwidth]{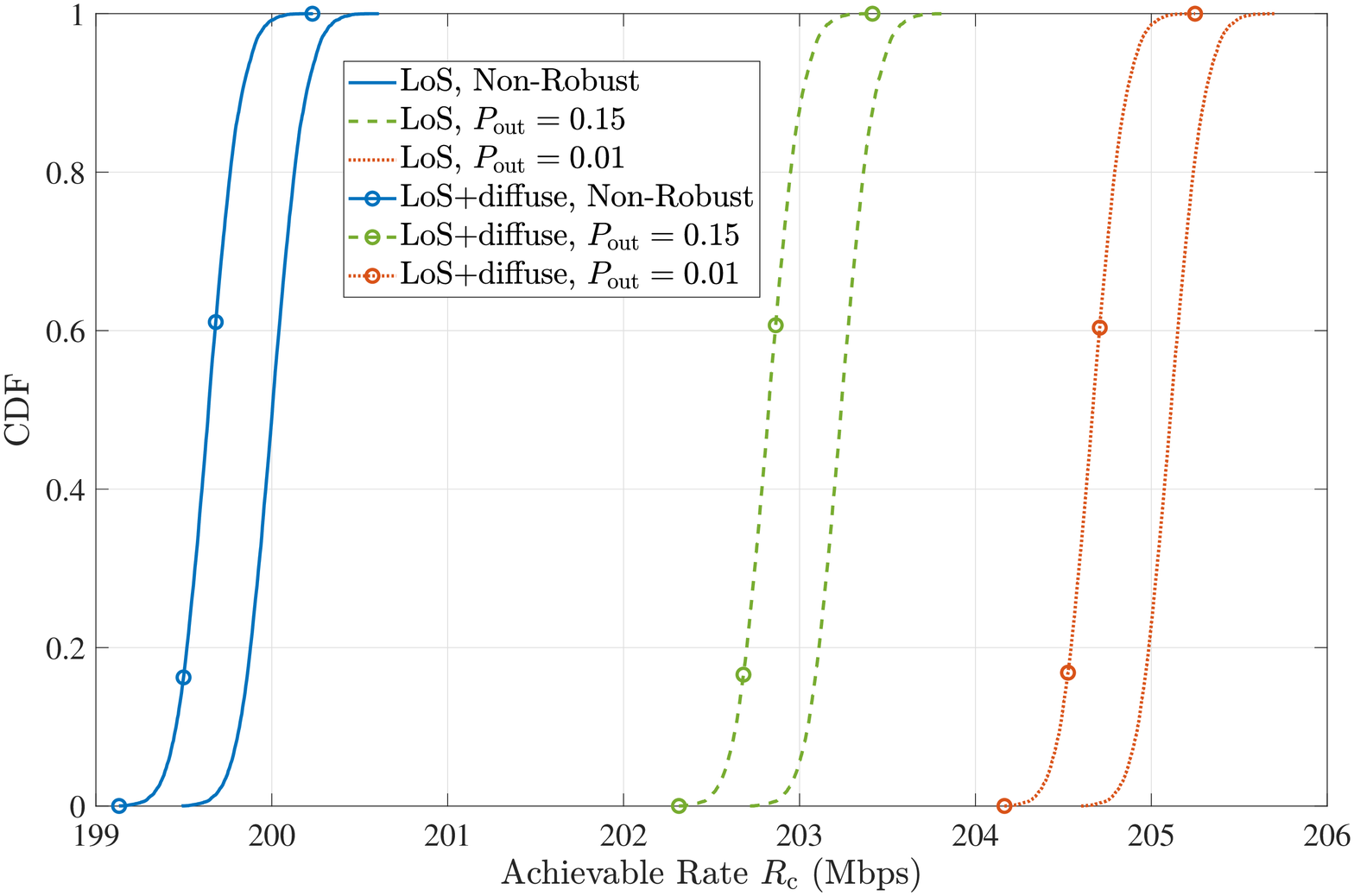}
        \vskip-0.2cm\centering {\footnotesize (a)}
    \end{minipage}
    \begin{minipage}[b]{0.35\textwidth}
        \centering
        \includegraphics[width =\textwidth]{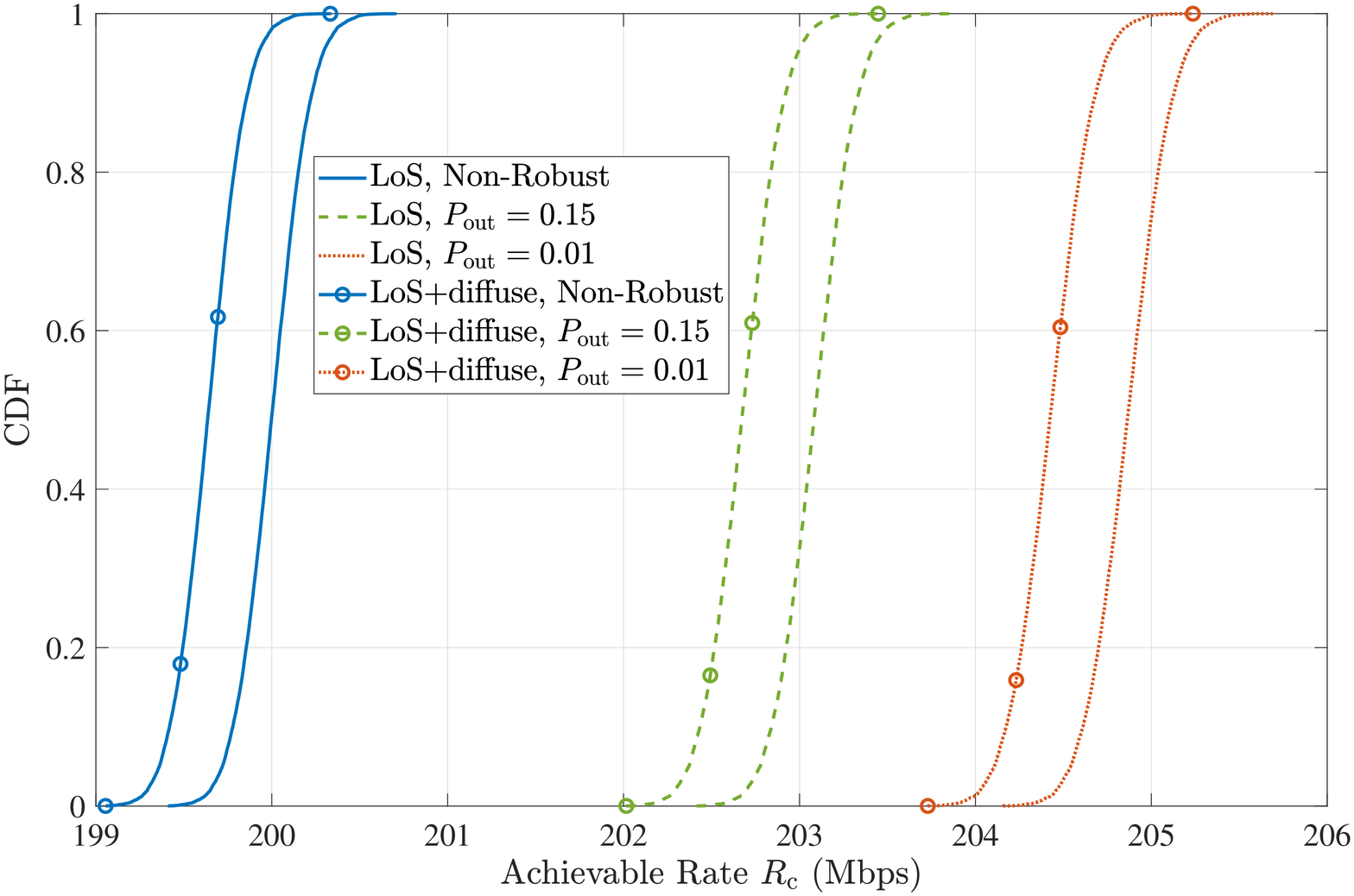}
        \vskip-0.2cm\centering {\footnotesize (b)}
    \end{minipage}
    \caption{CDF of the achievable rate $R_\mathrm{c}$ under the assumption of the Gaussian distributed positioning error $\mathbf{e_p}$ with rate thresholds $\bar{r}=200 \mathrm{Mbps}$ and different numbers of LEDs (a) $M = 3$; (b) $M = 6$.}
    \label{fig:cdf_rate_case1}
\end{figure}

\begin{figure}[htbp]
    \centering
    \begin{minipage}[b]{0.35\textwidth}
        \centering
        \includegraphics[width =\textwidth]{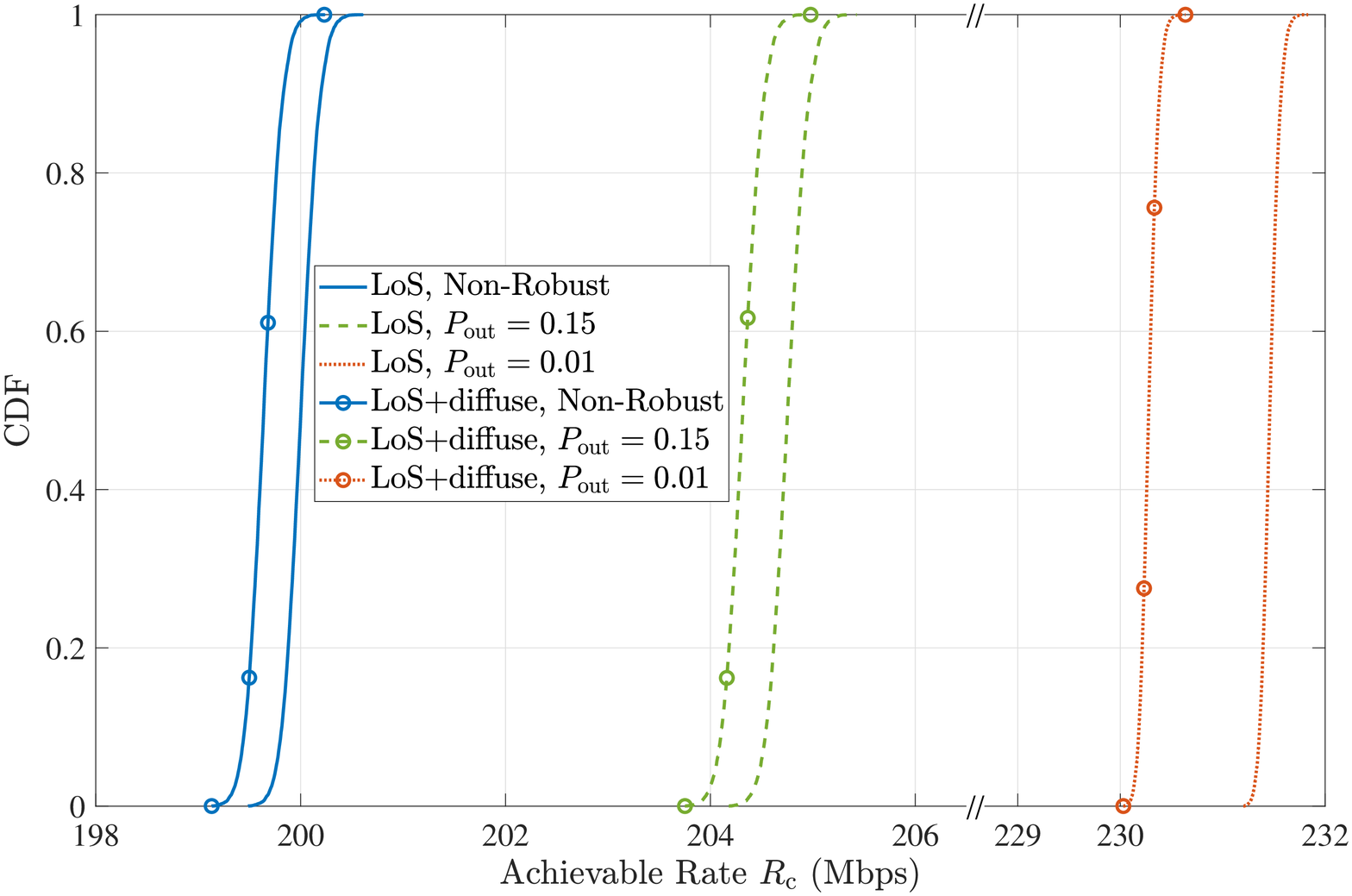}
        \vskip-0.2cm\centering {\footnotesize (a)}
    \end{minipage}
    \begin{minipage}[b]{0.35\textwidth}
        \centering
        \includegraphics[width =\textwidth]{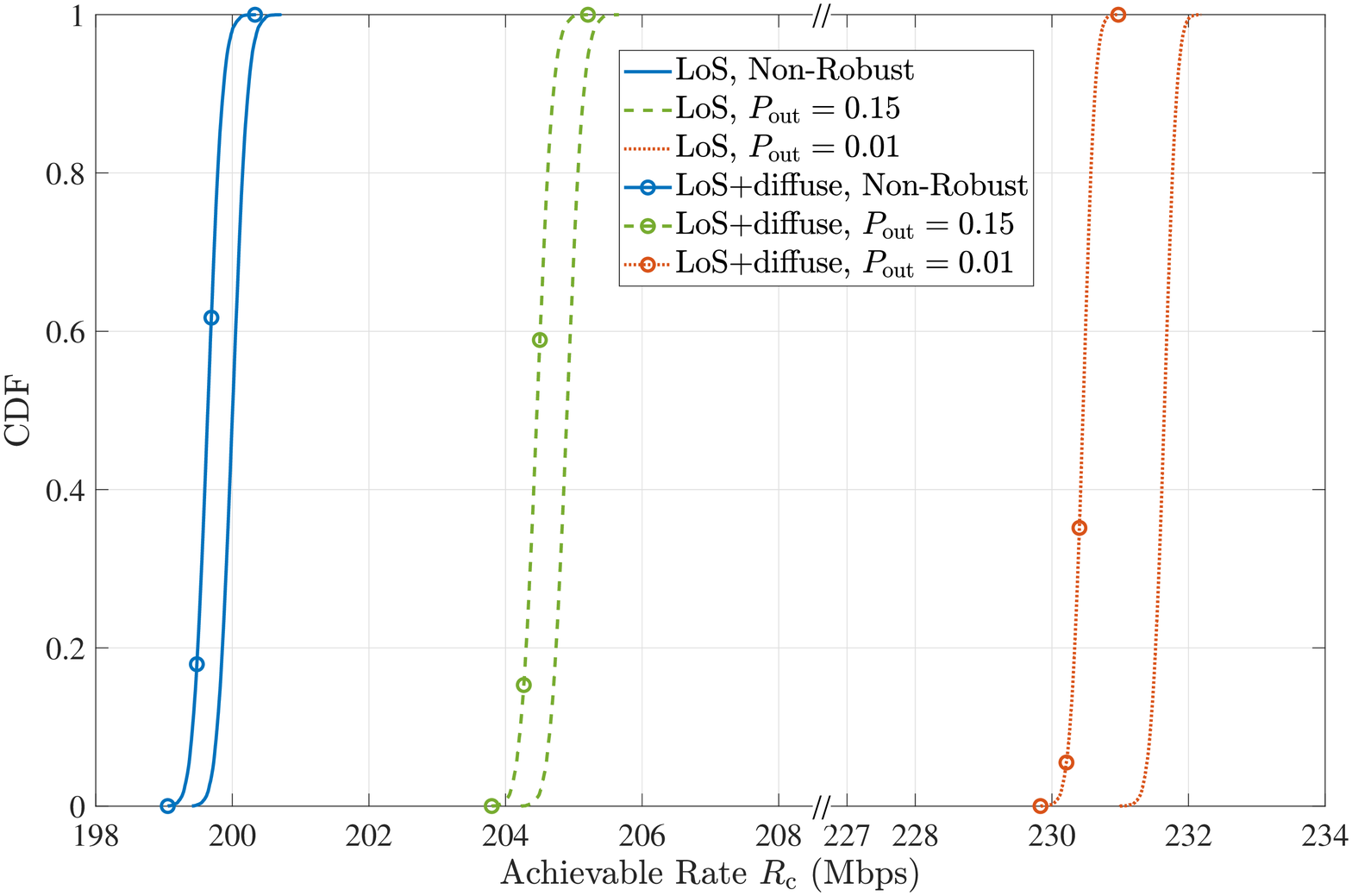}
        \vskip-0.2cm\centering {\footnotesize (b)}
    \end{minipage}
    \caption{CDF of the achievable rate $R_\mathrm{c}$ under the assumption of the arbitrary distributed positioning error $\mathbf{e_p}$ with rate thresholds $\bar{r}=200 \mathrm{Mbps}$ and different numbers of LEDs (a) $M = 3$; (b) $M = 6$.}
    \label{fig:cdf_rate_case2}
\end{figure}

We first evaluate the robust performance of the proposed power allocation schemes by Monte Carlo simulation, and the cumulative distribution functions (CDF) of the achievable data rate $R_{\mathrm{c}}$ are presented in Fig. \ref{fig:cdf_rate_case1} and Fig. \ref{fig:cdf_rate_case2}. In the initial stages, there is only a DC signal at each LED to provide illumination and positioning, i.e., $P_{\mathrm{c}}=0$, $P_{\mathrm{p},i}=\min\left\{P_{\mathrm{total}}\middle/M,\bar{P}_{\mathrm{p}}\right\}$. Then, the $10000$  random samples of  the positioning error $\mathbf{e}_{\mathrm{p}}$ are independently generated based on the distribution $\mathcal{N}\left(\mathbf{0},\mathbf{J}_{\mathbf{u}}^{-1}\left(\mathbf{p}_{\mathrm{p}}\right)\right)$. The estimated location $\hat{\mathbf{u}}$ can be derived by \eqref{e_p}. Finally, the result of power allocation is fed into a practical simulation environment to evaluate the achievable data rate $R_{\mathrm{c}}$. Besides, the mean of $\hat{\mathbf{u}}$ is considered as the replacement of the actual  location $\mathbf{u}$ in problems \eqref{VLPC_problem_perfect }, \eqref{Gauss_1}, and \eqref{VLPC_problem_arbitrary}.

Meanwhile, the proposed schemes are also  verified by both LOS and diffuse links (LOS+diffuse). According to the classical optical wireless channel model \cite{jungnickelPhysicalModelWireless2002}, the channel impulse response $h\left(t\right)$ is given by
    \begin{align}
        h\left(t\right)&=h_{i,\mathrm{Los}}\left(t\right)+h_{i,\mathrm{diffuse}}\left(t-\Delta T\right)\nonumber\\
        &=g_{i,\mathrm{LOS}}\delta\left(t\right)+\frac{\eta_{i,\mathrm{diffuse}}}{\tau}e^{-\frac{t-\Delta T}{\tau}}u\left(t-\Delta T\right),
    \end{align}
where $\eta_{i,\mathrm{diffuse}}$ is the power efficiency of the diffuse link, $\tau$ denotes the exponential decay time, $\Delta T$ is the delay between the LOS signal and the diffuse signal, $\delta\left(\cdot\right)$ is the Dirac function, and $u\left(\cdot\right)$ is the unit step function. We assume that $\left.{g_{i,\mathrm{LOS}}^2}\middle/{\eta_{i,\mathrm{diffuse}}^2}\right.=12\mathrm{dB}$, $\tau=15\mathrm{ns}$, and $\Delta T = 10\mathrm{ns}$. If the inter-symbol interference (ISI) due to the diffuse link is treated as noise, the achievable rate of the LOS+diffuse link is given by\cite{chenImpactMultipleReflections2020,maCapacityBoundsInterference2019}
\begin{align}
    R_{\mathrm{c}}\geq W\log_2\left(\frac{2\pi W\sigma_c^2+\left(P_1+P_2\right)e^{1+2\left(\alpha+\gamma\varepsilon\right)}}{2\pi W\sigma_c^2+2\pi\varepsilon P_2}\right),
\end{align}
where $P_1$ and $P_2$ denote the power without ISI and the power with ISI, respectively,
\begin{subequations}
    \begin{align}
        P_{1}&\triangleq \int_{0}^{\frac{1}{W}}\left|h\left(t\right)\right|^2P_{\mathrm{c}}\,\mathrm{d}t\\
        &= g_{i,\mathrm{LOS}}^2P_{\mathrm{c}}+\frac{\eta_{i,\mathrm{diffuse}}^2P_{\mathrm{c}}}{\tau^2}\int_{\Delta T}^{\frac{1}{W}}\exp\left(-\frac{2\left(t-\Delta T\right)}{\tau}\right)\,\mathrm{d}t\\
        &=g_{i,\mathrm{LOS}}^2P_{\mathrm{c}}+\frac{\eta_{i,\mathrm{diffuse}}^2P_{\mathrm{c}}}{2\tau}\left(1-\exp\left(\frac{2\Delta TW-2}{W\tau}\right)\right),\\
        P_{2}&\triangleq\int_{\frac{1}{W}}^{\infty}\left|h\left(t\right)\right|^2P_{\mathrm{c}}\,\mathrm{d}t\\
        &=\frac{\eta_{i,\mathrm{diffuse}}^2P_{\mathrm{c}}}{\tau^2}\int_{\frac{1}{W}}^{\infty}\exp\left(-\frac{2\left(t-\Delta T\right)}{\tau}\right)\,\mathrm{d}t\\
        &=\frac{\eta_{i,\mathrm{diffuse}}^2P_{\mathrm{c}}}{2\tau}\exp\left(\frac{2\Delta TW-2}{W\tau}\right).
    \end{align}
\end{subequations}

Fig. \ref{fig:cdf_rate_case1}(a) illustrates the CDF of the achievable rate $R_{\mathrm{c}}$ under the assumption of the Gaussian distributed positioning error $\mathbf{e_p}$ with $M=3$ LEDs, and different maximum tolerable outage probabilities $P_{\mathrm{out}}=0.01$, $P_{\mathrm{out}}=0.15$ and nonrobust ($P_{\mathrm{out}}\to\infty$) for only the LOS link and the LOS+diffuse link, respectively. For the nonrobust method, the outage probability with only the LOS link is about $0.5$, and it is close to $1$ for the LOS+diffuse case, which significantly exceeds the maximum tolerated outage probability requirement. On the other hand, the outage probability of the proposed robust power allocation in Section \ref{sec:robust-VLPC-design-with-gaussian-distributed-bfermp} is close to $0$, even for the LOS+diffuse case, which is well below the requirements $P_{\mathrm{out}}=0.01$ and $P_{\mathrm{out}}=0.15$.
Fig. \ref{fig:cdf_rate_case1}(b) depicts the same results for $6$ LEDs.
For the same $P_{\text{out}}$ and channel type, the CDF curves in Fig. \ref{fig:cdf_rate_case1}(b) are lower than those in Fig. \ref{fig:cdf_rate_case1}(a).
In other words, the conservatism of the proposed robust scheme can be reduced slightly as the number of LEDs increases. In addition, if the other settings are the same, the achievable rate with $P_{\text{out}}=0.01$ is highest, and the rate of the nonrobust case is lowest. Thus, it leads to a higher actual rate to guarantee stricter outage probability constraint.

Fig. \ref{fig:cdf_rate_case2} shows the case of the arbitrary positioning error distribution, and the chance constraint of the achievable rate is also satisfied. Comparing with Fig. \ref{fig:cdf_rate_case1}, the CVaR-based scheme provides a higher communication rate than the Bernstein-based scheme for the same scenario. Thus, the CVaR-based scheme is more robust than the scheme based on the Bernstein-type inequality, because of less utilized prior information. On the other hand, the CVaR-based scheme can be applied widely in practical scenarios.

\subsection{CRLB Versus the Rate Threshold}\label{sec:simulation_crlb_rate}

\begin{figure}[htbp]
    \centering
    \begin{minipage}[b]{0.35\textwidth}
        \centering
        \includegraphics[width =\textwidth]{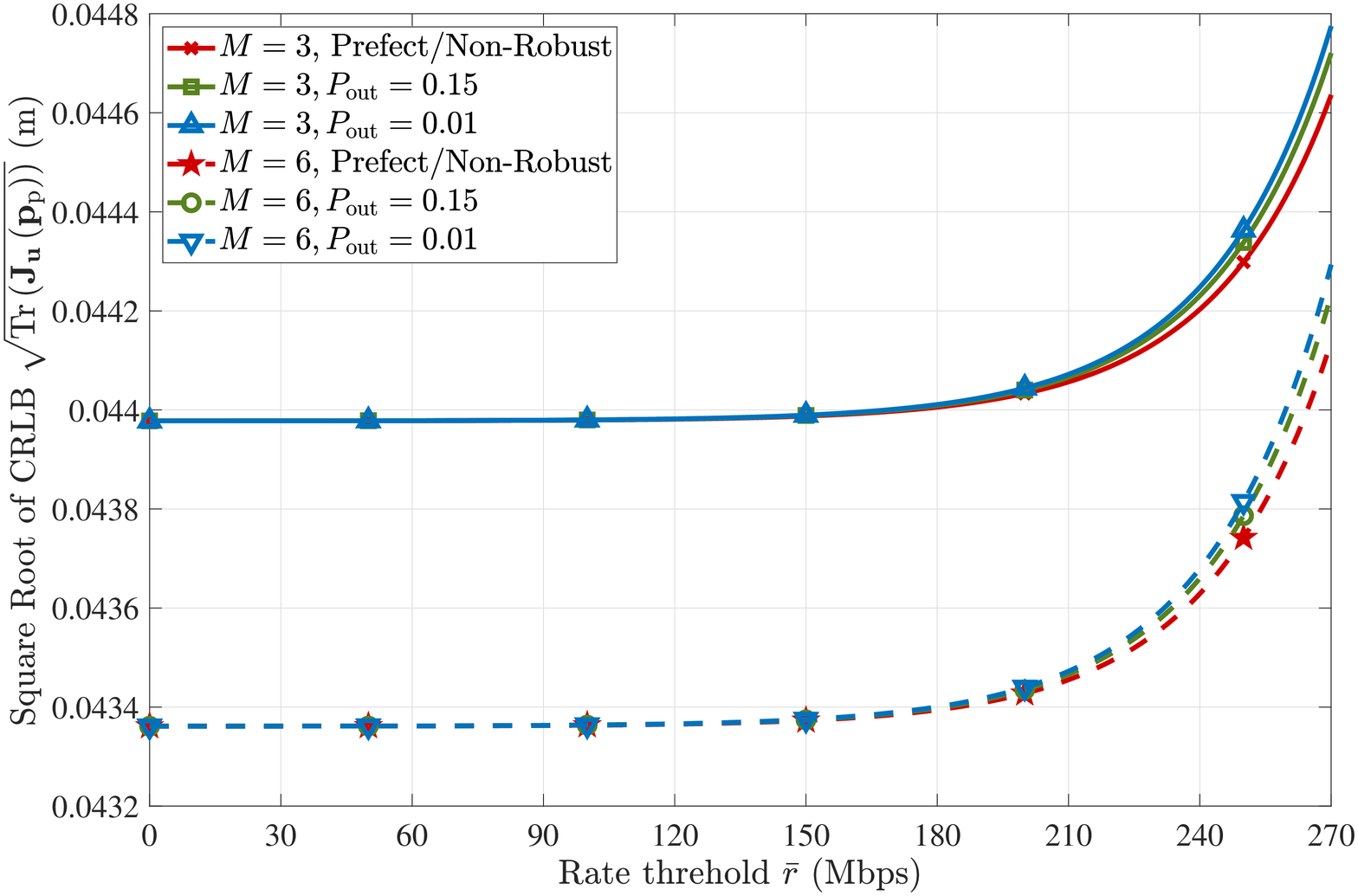}
        \vskip-0.2cm\centering {\footnotesize (a)}
    \end{minipage}
    \begin{minipage}[b]{0.35\textwidth}
        \centering
        \includegraphics[width =\textwidth]{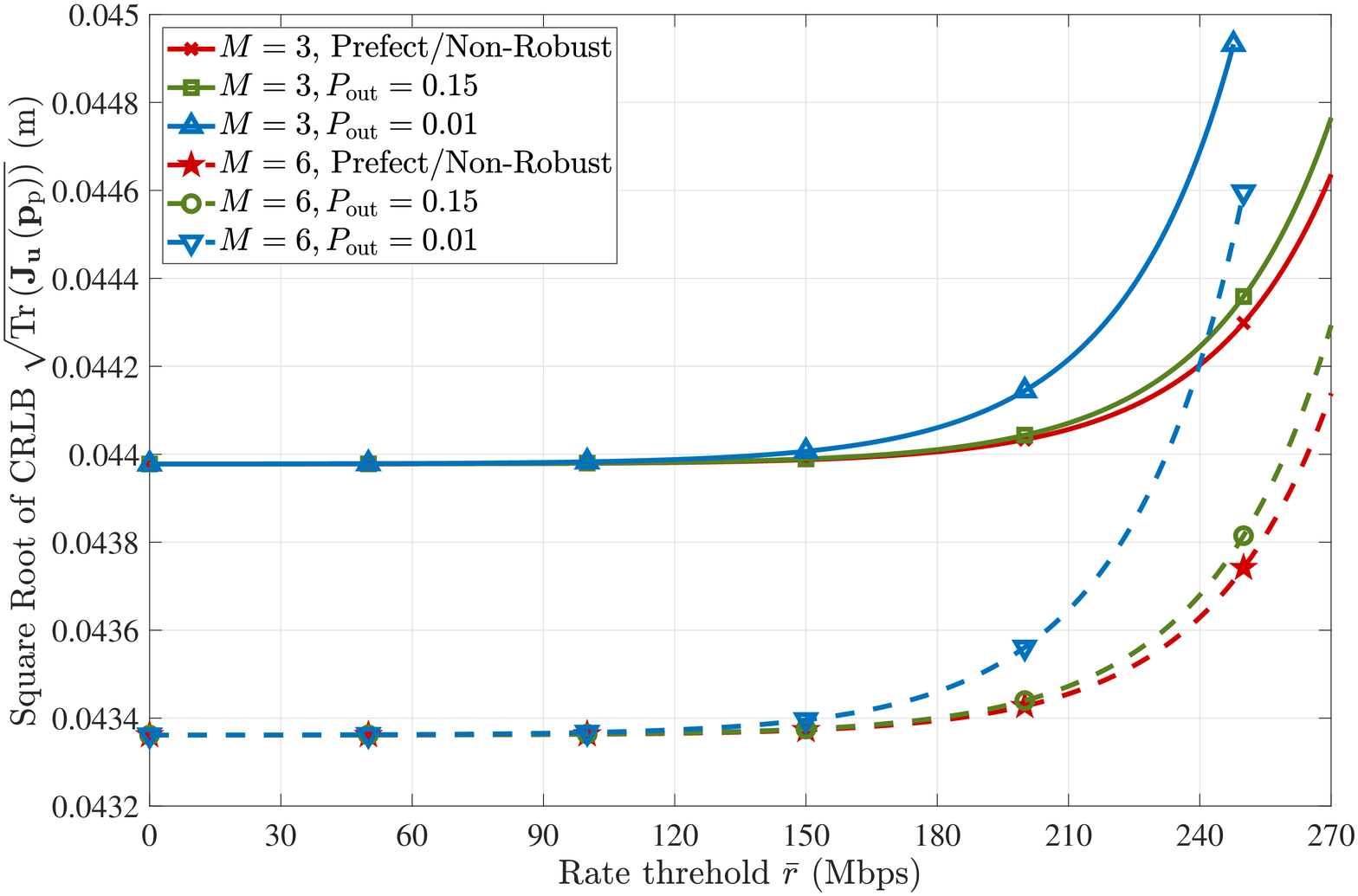}
        \vskip-0.2cm\centering {\footnotesize (b)}
    \end{minipage}
    \caption{Square root of the CRLB versus rate thresholds $\bar{r}$ with the fixed UE's location estimation $\hat{\mathbf{u}}=\mathbf{u}$, only the LOS link, under the different positioning error distribution assumptions (a) Gaussian distribution; (b) Arbitrary distribution.}
    \label{fig:crlb_rate}
\end{figure}

In this subsection, we show the relationship between the positioning error and the communication requirement for the nonrobust and two proposed robust schemes. The square root of the CRLB versus the achievable rate threshold $\bar{r}$ is investigated in Fig. \ref{fig:crlb_rate}(a) and Fig. \ref{fig:crlb_rate}(b) under the Gaussian and arbitrary positioning error distributed assumptions, respectively. To represent the average performance, it is assumed that there is only the LOS channel, and the UE's location estimation is the actual position, i.e., $\hat{\mathbf{u}}=\mathbf{u}$.

Fig. \ref{fig:crlb_rate}(a) shows the results under the Gaussian distributed assumption with $M=3, 6$ LEDs, and the maximum tolerable outage probabilities are $P_{\text{out}}=0.01$, $P_{\text{out}}=0.15$ and nonrobust case. Since $\hat{\mathbf{u}}=\mathbf{u}$, the nonrobust case is equivalent to the perfect  positioning case. We observe that the positioning performance degrades as the minimum rate requirement increases, which means that there is a tradeoff between the communication performance and the positioning precision. When the LED number is fixed, and the rate threshold $\bar{r}$ is low enough, such as $\bar{r}\leq 150 \text{Mbps}$,  the CRLB changes slowly. However, if $\bar{r}$ is high enough, it degrades rapidly. Besides, for the same rate threshold and the LED number, stricter $P_{\text{out}}$ will lead to worse CRLB. On the other hand, the CRLB can be obviously reduced as  the  LED number increases from $M=3$ to $M=6$.

Comparing with Fig. \ref{fig:crlb_rate}(a), Fig. \ref{fig:crlb_rate}(b) is set to the same scenario except for the distribution assumption, and there are similar trends of the CRLB versus the rate for different parameters. However, the CVaR-based scheme leads to more drastic positioning performance loss versus the rate threshold $\bar{r}$ and the outage probabilities threshold $P_{\text{out}}$.

\subsection{Power Allocation Versus the Rate Threshold}

\begin{figure}[htbp]
    \centering
    \begin{minipage}[b]{0.35\textwidth}
        \centering
        \includegraphics[width =\textwidth]{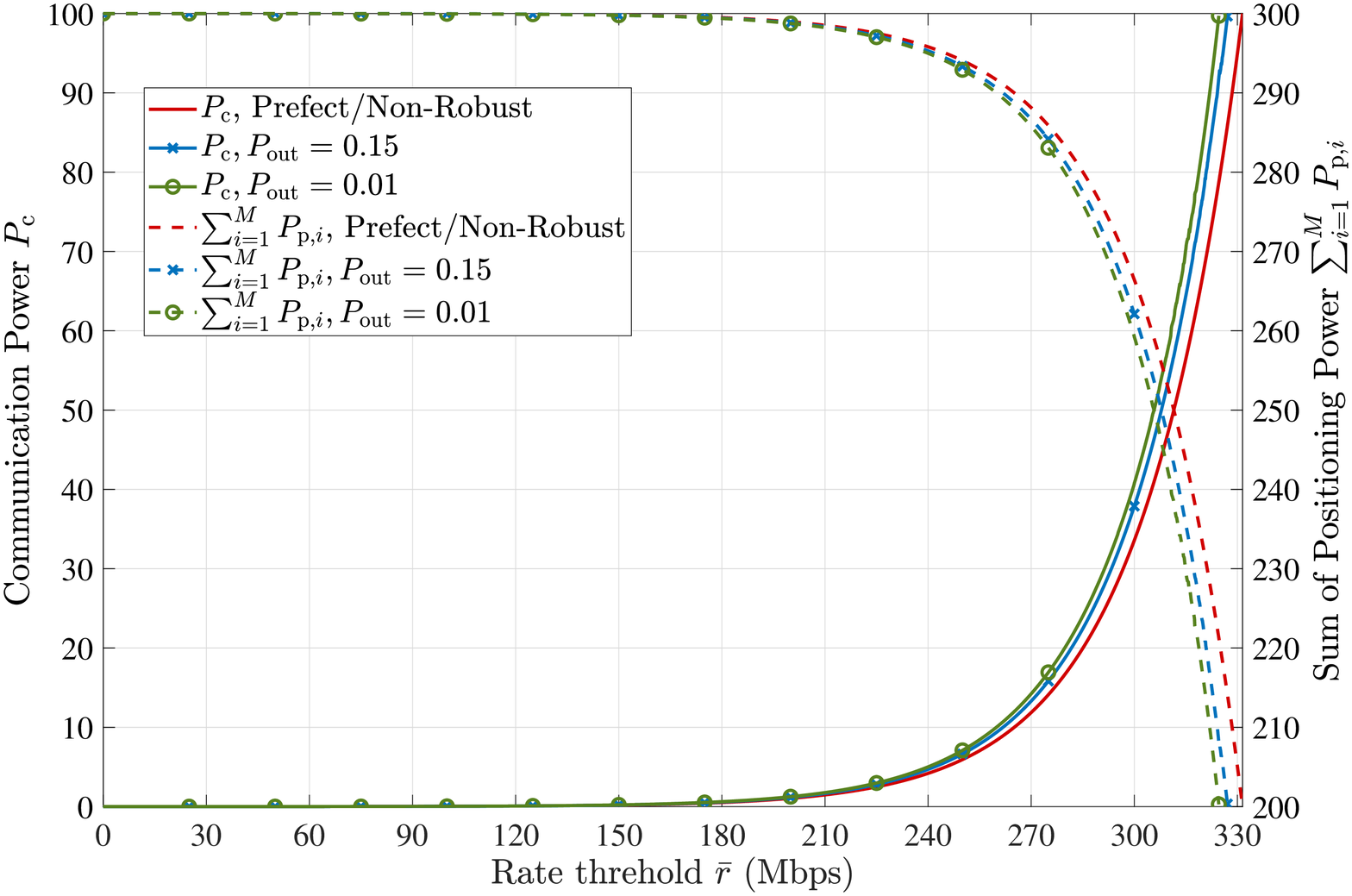}
        \vskip-0.2cm\centering {\footnotesize (a)}
    \end{minipage}
    \begin{minipage}[b]{0.35\textwidth}
        \centering
        \includegraphics[width =\textwidth]{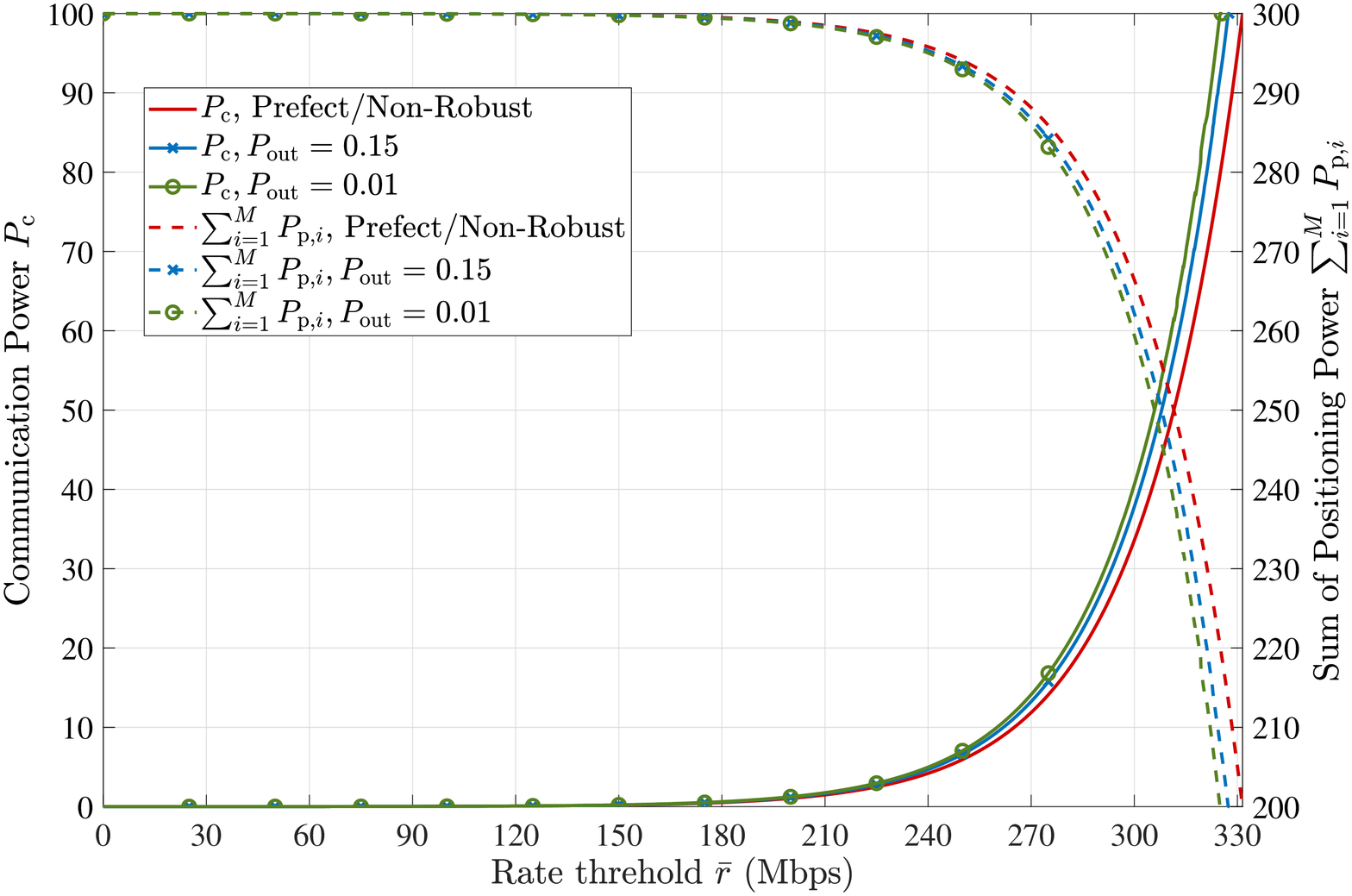}
        \vskip-0.2cm\centering {\footnotesize (b)}
    \end{minipage}
    \caption{Power allocation $\mathbf{p}_\mathrm{p}$ and $P_\mathrm{c}$  versus the rate threshold $\bar{r}$  under the assumption of the Gaussian distributed positioning error $\mathbf{e_p}$  with the fixed UE's location estimation $\hat{\mathbf{u}}=\mathbf{u}$, only the LOS link, and different numbers of LEDs (a) $M = 3$; (b) $M = 6$.}
    \label{fig:power_rate_case1}
\end{figure}

\begin{figure}[htbp]
    \centering
    \begin{minipage}[b]{0.35\textwidth}
        \centering
        \includegraphics[width =\textwidth]{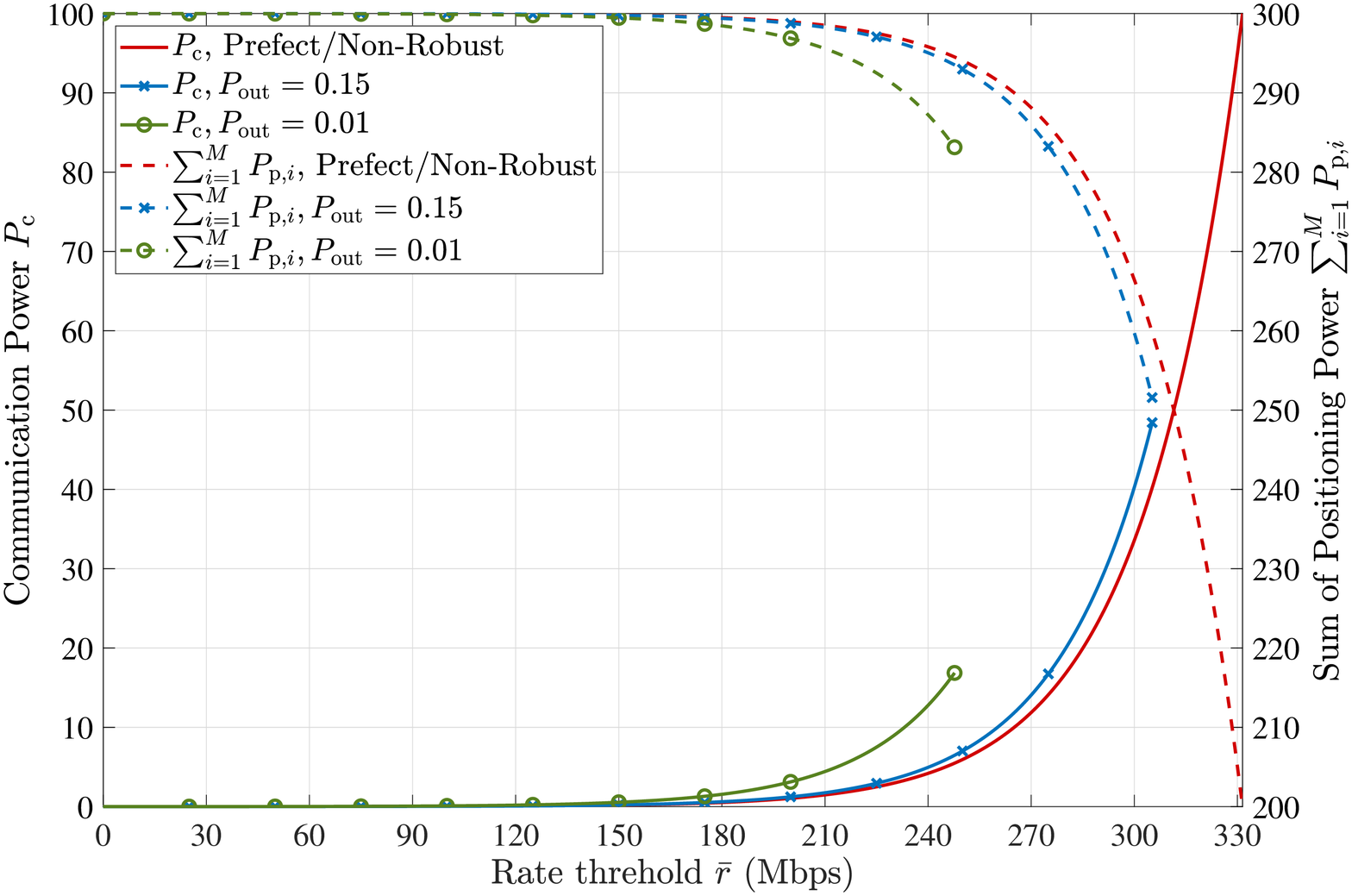}
        \vskip-0.2cm\centering {\footnotesize (a)}
    \end{minipage}
    \begin{minipage}[b]{0.35\textwidth}
        \centering
        \includegraphics[width =\textwidth]{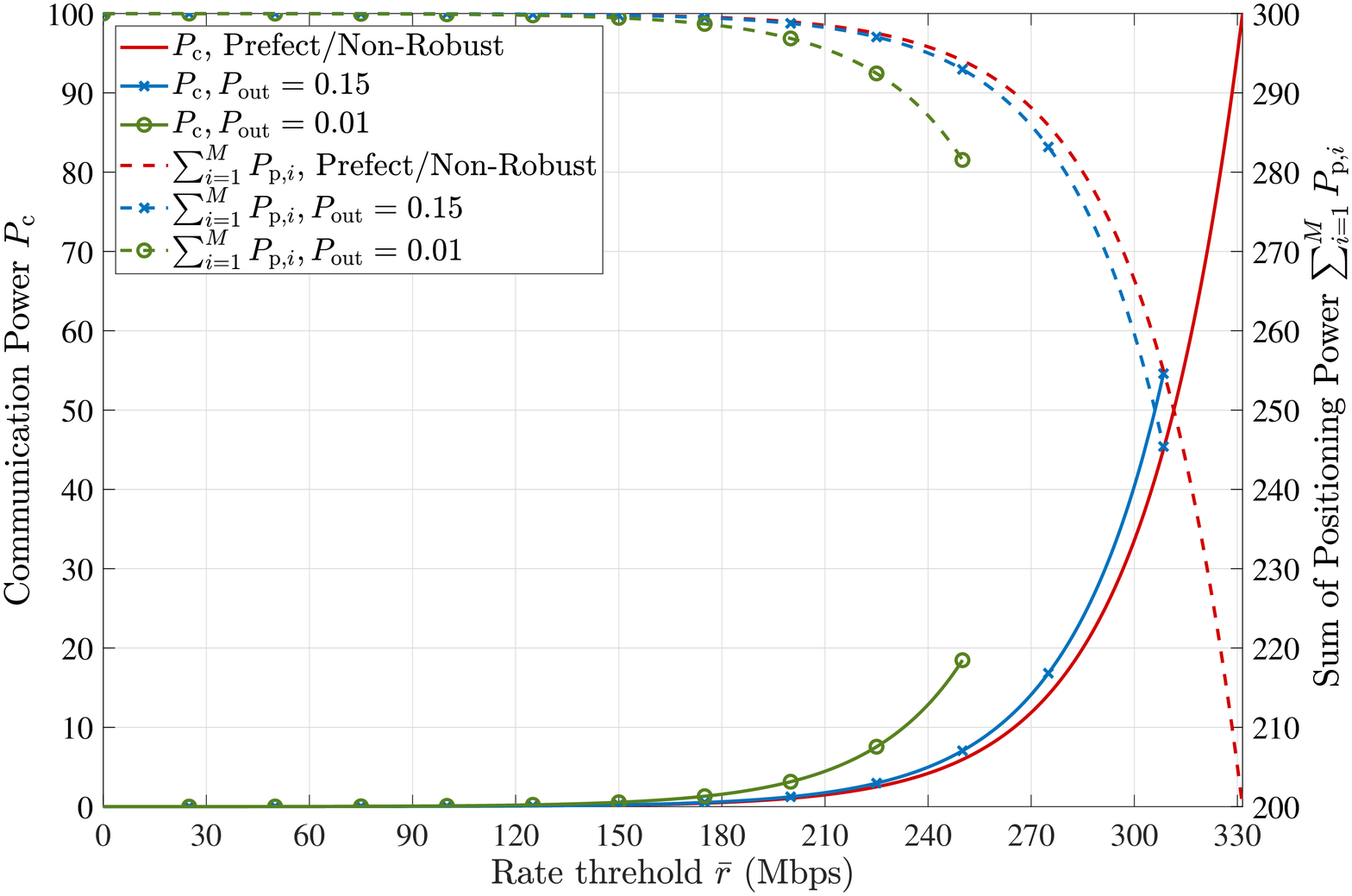}
        \vskip-0.2cm\centering {\footnotesize (b)}
    \end{minipage}
    \caption{Power allocation $\mathbf{p}_\mathrm{p}$ and $P_\mathrm{c}$ versus rate thresholds $\bar{r}$  under the assumption of the arbitrary distributed positioning error $\mathbf{e_p}$  with the fixed UE's location estimation $\hat{\mathbf{u}}=\mathbf{u}$, only the LOS link, and different numbers of LEDs (a) $M = 3$; (b) $M = 6$.}
    \label{fig:power_rate_case2}
\end{figure}

To further derive the performance tradeoff between communication and positioning, the power allocation results versus the rate threshold are illustrated in Fig. \ref{fig:power_rate_case1} and Fig. \ref{fig:power_rate_case2}, assuming Gaussian and arbitrary distributed positioning error, respectively. Other design parameters are the same as those in Section \ref{sec:simulation_crlb_rate}.

From Fig. \ref{fig:power_rate_case1} and Fig. \ref{fig:power_rate_case2}, the allocated communication power $P_{\text{c}}^*$ increases as the minimum rate requirement $\bar{r}$ increases, and the rate of change also increases continuously. To minimize the CRLB, the remaining power is always allocated to the positioning signals as much as possible within the constraints. For problems \eqref{VLPC_problem_perfect }, \eqref{VLPC_problem}, and \eqref{VLPC_problem_arbitrary}, the allocated positioning power $\sum_{i=1}^{M}P_{\text{p},i}^{*}$ should satisfy $\min\bigl\{P_{\text{total}}-P_{\text{c}}^*,M\bar{P}_{\text{p}}\bigr\}$. Due to the limited total power, $\sum_{i=1}^{M}P_{\text{p},i}^{*}$ will decrease as the minimum rate requirement $\bar{r}$ increases, which leads to the CRLB degradation.


In addition, more power will be allocated to VLC, if a stricter outage probability constraint or a more robust power allocation scheme is adopted. This causes the sort of achievable rates in Fig. \ref{fig:cdf_rate_case1} and Fig. \ref{fig:cdf_rate_case2}. When other settings are the same, the achievable rate with $P_{\text{out}}=0.01$ is the highest, and the rate of the nonrobust case is the lowest. Meanwhile, the CVaR-based scheme provides a higher communication rate than the Bernstein-based scheme for the same scenario. This is also the reason for the variation of the CRLB, as shown in Fig. \ref{fig:crlb_rate} with different parameters and schemes.

According to \eqref{delta_g1}, \eqref{VLPC_problem_e_p} and \eqref{VLPC_problem_arbitrary_e_p_1}, the positioning error distribution can directly influence the distribution of the channel estimation error. Although the allocated positioning power with $M=3$ LEDs approximates the power with $M=6$ LEDs in both Fig. \ref{fig:power_rate_case1} and Fig. \ref{fig:power_rate_case2}, the more LEDs can improve the positioning performance, as illustrated in Fig. \ref{fig:crlb_rate}. Thus, it can reduce the variance of the channel estimation error to increase the  LED number. For communication, the conservatism of the robust schemes can be mitigated, as shown in Fig. \ref{fig:cdf_rate_case1} and Fig. \ref{fig:cdf_rate_case2}.

\subsection{Cumulative Distribution Functions of Positioning Errors}

\begin{figure}[htbp]
    \centering
    \begin{minipage}[b]{0.35\textwidth}
            \centering
            \includegraphics[width =\textwidth]{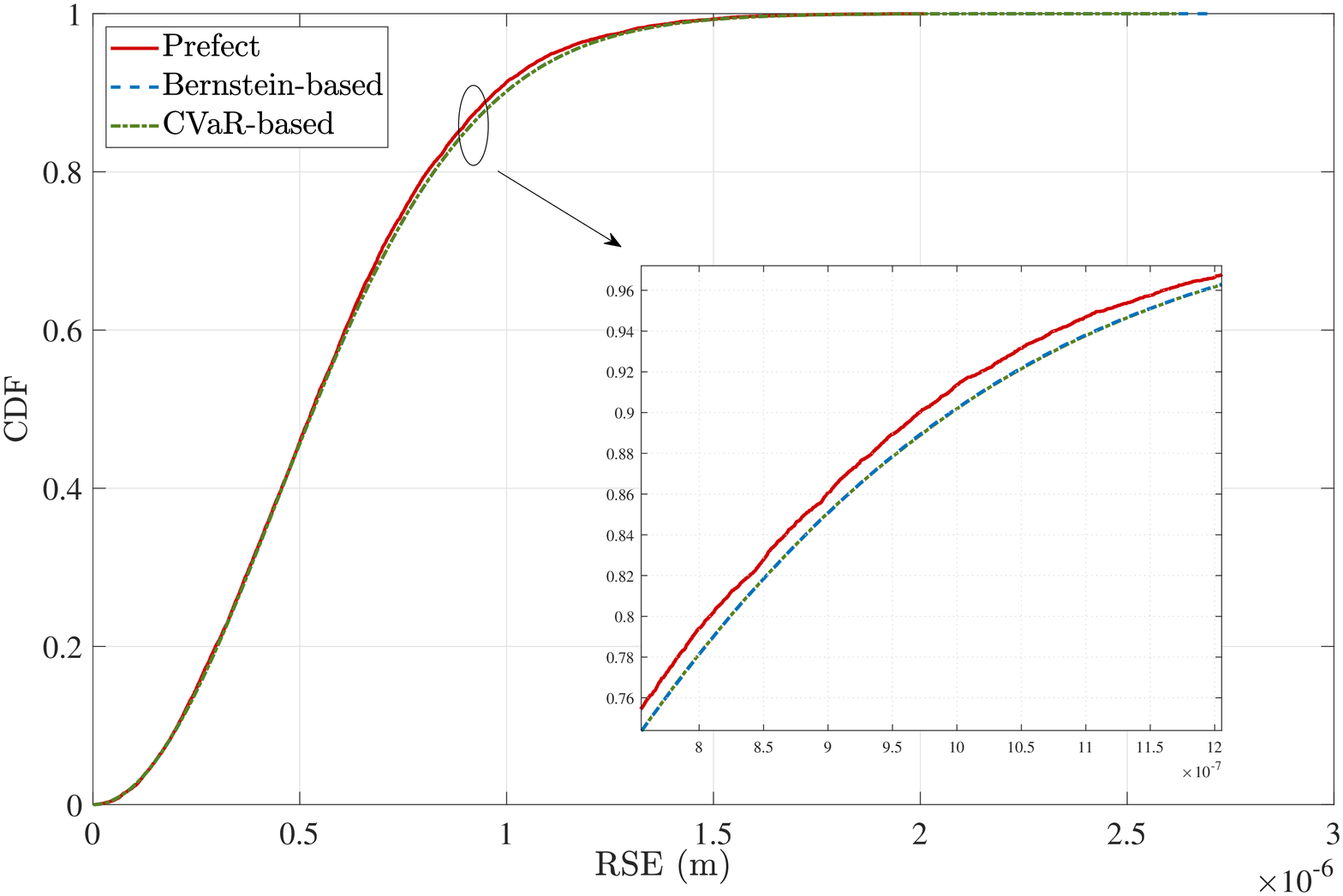}
            \vskip-0.2cm\centering {\footnotesize (a)}
        \end{minipage}
    \begin{minipage}[b]{0.35\textwidth}
            \centering
            \includegraphics[width =\textwidth]{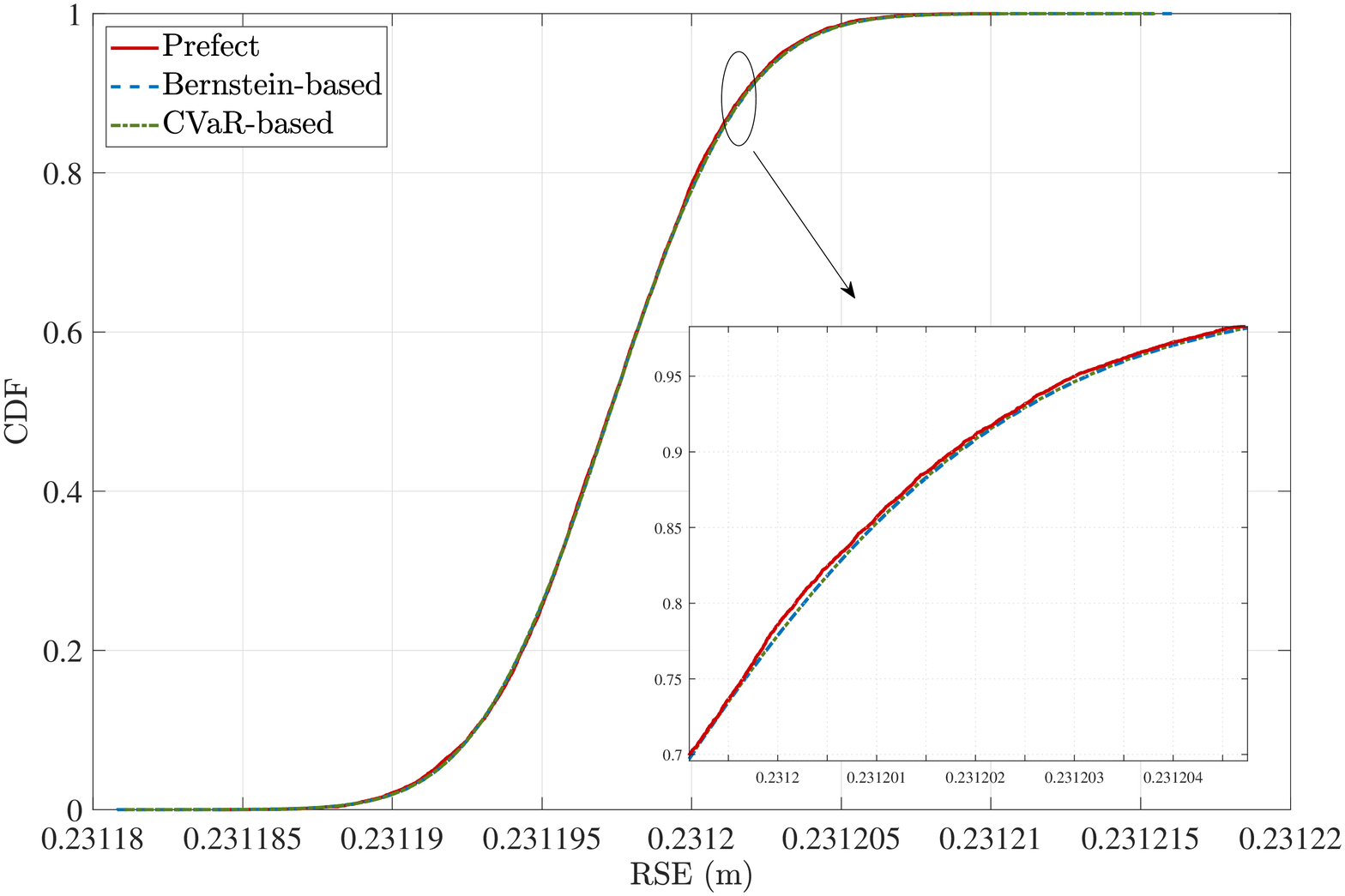}
            \vskip-0.2cm\centering {\footnotesize (b)}
        \end{minipage}
    \caption{CDF of RSE for the perfect  power allocation, Bernstein-based power allocation, and CVaR-based power allocation with $M=3$ LEDs, rate thresholds $\bar{r}=200 \mathrm{Mbps}$, maximum tolerable outage probability $P_{\mathrm{out}}=0.01$, and different channels (a) only LOS; (b) LOS+diffuse.}
    \label{fig:cdf_pos_err}
\end{figure}

To evaluate the positioning performance loss of robust VLPC schemes, the RSS-based triangulation process was simulated with only the LOS link and the LOS+diffuse link. The location is  estimated using the nonlinear least squares method. Besides, the positioning error is calculated using the root-square error (RSE). The CDFs of positioning errors are shown in Fig. \ref{fig:cdf_pos_err}.

From both Fig. \ref{fig:cdf_pos_err}(a) and Fig. \ref{fig:cdf_pos_err}(b), two robust VLPC power allocation schemes lead to only a slight loss of positioning precision. Thus, the proposed robust power allocation schemes are also effective for VLP.

\section{conclusion}\label{sec:conclusion}

In this work, we investigated the inherent coupling between VLP and VLC through the relationship between the optical channel and the location.
In other words, channel estimation can be implemented using the positioning result.
After deriving the CRLB for VLP and the achievable rate for VLC, we unveiled the tradeoff between VLP and VLC by the relationship between the channel estimation error and the positioning error.
Furthermore, we proposed two robust power allocation schemes for VLPC to minimize the CRLB under the power constraints and QoS requirements, where the positioning error distributions are assumed to follow either the Gaussian distribution or an arbitrary distribution with a known mean and variance. Under the Gaussian distributed assumption, the Bernstein-type inequality is utilized to tackle the communication rate outage constraints, and the problem was converted into a stricter convex SDP by exploiting the matrix norm. For the arbitrary distributed case, the problem was approximated by a more tractable form through the worst-case CVaR and SCA based on the first-order Taylor expansion.
Finally, our simulation results demonstrated the effectiveness of our two proposed robust VLPC power allocation schemes for both localization and communications.

  \begin{appendices}
   \section{Derivation of the formulation \eqref{FIM1}}\label{sec:derivation-of-the-formulation-eqreffim1}

The derivative of the log-likelihood function $\Lambda \left( \bm{\mathbf{u}}  \right)$ with respect to $\bm{\mathbf{u}}$ is given by\cite{Kay_1993}
\begin{align}
\frac{{\partial \Lambda \left( \bm{\mathbf{u}}  \right)}}{{\partial {x_{\mathrm{u}}}}} &=  - \frac{1}{{\sigma _{\mathrm{p}}^2}}\sum\limits_{i\in\tilde{\mathcal{M}}} \int_0^{{T_{\mathrm{p}}}} \left( {g_i}\frac{{\partial {g_i}}}{{\partial {x_{\mathrm{u}}}}}{{{ {{P_{{\mathrm{p}},i}}} {s^2_{{\mathrm{p,}}i}\left(t\right)}}}}\right.\nonumber\\
&\qquad\qquad\quad\left.- {y_{{\mathrm{p}},i}\left(t\right)}\frac{{\partial {g_i}}} {{\partial {x_{\mathrm{u}}}}}{\sqrt {{P_{{\mathrm{p}},i}}} {s_{{\mathrm{p,}}i}\left(t\right)}} \right)\,\mathrm{d}t, \\
\frac{{\partial \Lambda \left( \bm{\mathbf{u}}  \right)}}{{\partial {y_{\mathrm{u}}}}} &=  - \frac{1}{{\sigma _{\mathrm{p}}^2}}\sum\limits_{i\in\tilde{\mathcal{M}}} \int_0^{{T_{\mathrm{p}}}} \left( {g_i}\frac{{\partial {g_i}}}{{\partial {y_{\mathrm{u}}}}}{{{ {{P_{{\mathrm{p}},i}}} {s^2_{{\mathrm{p,}}i}\left(t\right)}}}}\right.\nonumber\\
&\qquad\qquad\quad\left. - {y_{{\mathrm{p}},i}\left(t\right)}\frac{{\partial {g_i}}} {{\partial {y_{\mathrm{u}}}}}{\sqrt {{P_{{\mathrm{p}},i}}} {s_{{\mathrm{p,}}i}\left(t\right)}} \right)\,\mathrm{d}t,
\end{align}
where $\frac{{\partial {g_i}}}{{\partial {x_{\mathrm{u}}}}}$ and $\frac{{\partial {g_i}}}{{\partial {y_{\mathrm{u}}}}}$ can be expressed as
     \begin{align}
\frac{{\partial {g_i}}}
{{\partial {x_{\mathrm{u}}}}} =  - \frac{\left( {m + 3} \right)\mu {\left( {{z_i} - {z_{\mathrm{u}}}} \right)^{m + 1}}\left( {{x_{\mathrm{u}}} - {x_i}} \right)}{{\left\| {{\mathbf{u}} - {{\mathbf{v}}_i}} \right\|^{{m + 5}}}},\\
\frac{{\partial {g_i}}}
{{\partial {y_{\mathrm{u}}}}} =  - \frac{\left( {m + 3} \right)\mu {\left( {{z_i} - {z_{\mathrm{u}}}} \right)^{m + 1}}\left( {{y_{\mathrm{u}}} - {y_i}} \right)}{{\left\| {{\mathbf{u}} - {{\mathbf{v}}_i}} \right\|^{{m + 5}}}},
   \end{align}

Furthermore, the  FIM  can be denoted as an explicit function between the position power ${{\mathbf{p}}_{\mathrm{p}}} = {\left[ {{P_{{\mathrm{p}},1}}, \ldots,{P_{{\mathrm{p}},M}}} \right]^T}$ and the unknown UE location ${\bm{\mathbf{u} }}$ in the 2D plane.
\begin{align}\label{fim}
{\mathbf{J}_{\bm{\mathbf{u} }}}\left( {{{\mathbf{p}}_{\mathrm{p}}}} \right) = \left[ {\begin{array}{*{20}{c}}
   {-\mathbb{E}\left( {\frac{{{\partial ^2}\Lambda \left( \bm{\mathbf{u}}  \right)}}
{{\partial {x_{\mathrm{u}}}\partial {x_{\mathrm{u}}}}}} \right)} & {-\mathbb{E}\left( {\frac{{{\partial ^2}\Lambda \left( \bm{\mathbf{u}}  \right)}}
{{\partial {x_{\mathrm{u}}}\partial {y_{\mathrm{u}}}}}} \right)}  \\
   {-\mathbb{E}\left( {\frac{{{\partial ^2}\Lambda \left( \bm{\mathbf{u}}  \right)}}
{{\partial {y_{\mathrm{u}}}\partial {x_{\mathrm{u}}}}}} \right)} & {-\mathbb{E}\left( {\frac{{{\partial ^2}\Lambda \left( \bm{\mathbf{u}}  \right)}}
{{\partial {y_{\mathrm{u}}}\partial {y_{\mathrm{u}}}}}} \right)}  \\
 \end{array} } \right],
\end{align}
where the elements of \eqref{fim} are given as
\begin{align}
\mathbb{E}\left( {\frac{{{\partial ^2}\Lambda \left( \bm{\mathbf{u}}  \right)}}{{\partial {x_{\mathrm{u}}}\partial {x_{\mathrm{u}}}}}} \right) &=  - \frac{\sum\limits_{i\in\tilde{\mathcal{M}}} P_{\mathrm{p},i}\mathbb{E}\left\{\int_0^{T_{\mathrm{p}}} s^2_{\mathrm{p},i}\left(t\right)\,\mathrm{d}t\right\} \frac{{\partial {g_i}}}{{\partial {x_{\mathrm{u}}}}}\frac{{\partial {g_i}}}{{\partial {x_{\mathrm{u}}}}}}{{\sigma _{\mathrm{p}}^2}},\\
\mathbb{E}\left( {\frac{{{\partial ^2}\Lambda \left( \bm{\mathbf{u}}  \right)}}{{\partial {y_{\mathrm{u}}}\partial {y_{\mathrm{u}}}}}} \right) &=  - \frac{\sum\limits_{i\in\tilde{\mathcal{M}}} P_{\mathrm{p},i}\mathbb{E}\left\{\int_0^{T_{\mathrm{p}}} s^2_{\mathrm{p},i}\left(t\right)\,\mathrm{d}t\right\} \frac{{\partial {g_i}}}{{\partial {y_{\mathrm{u}}}}}\frac{{\partial {g_i}}}{{\partial {y_{\mathrm{u}}}}}}{{\sigma _{\mathrm{p}}^2}},\\
\mathbb{E}\left( {\frac{{{\partial ^2}\Lambda \left( \bm{\mathbf{u}}  \right)}}{{\partial {x_{\mathrm{u}}}\partial {y_{\mathrm{u}}}}}} \right) &=  - \frac{\sum\limits_{i\in\tilde{\mathcal{M}}} P_{\mathrm{p},i}\mathbb{E}\left\{\int_0^{T_{\mathrm{p}}} s^2_{\mathrm{p},i}\left(t\right)\,\mathrm{d}t\right\} \frac{{\partial {g_i}}}{{\partial {y_{\mathrm{u}}}}}\frac{{\partial {g_i}}}{{\partial {x_{\mathrm{u}}}}}}{{\sigma _{\mathrm{p}}^2}},\\
\mathbb{E}\left( {\frac{{{\partial ^2}\Lambda \left( \bm{\mathbf{u}}  \right)}}{{\partial {y_{\mathrm{u}}}\partial {x_{\mathrm{u}}}}}} \right) &=  - \frac{\sum\limits_{i\in\tilde{\mathcal{M}}} P_{\mathrm{p},i}\mathbb{E}\left\{\int_0^{T_{\mathrm{p}}} s^2_{\mathrm{p},i}\left(t\right)\,\mathrm{d}t\right\} \frac{{\partial {g_i}}}{{\partial {x_{\mathrm{u}}}}}\frac{{\partial {g_i}}}{{\partial {y_{\mathrm{u}}}}}}{{\sigma _{\mathrm{p}}^2}}.
\end{align}

According to $\mathbb{E}\left\{\int_0^{T_{\mathrm{p}}} s^2_{\mathrm{p},i}\left(t\right)\,\mathrm{d}t\right\} = \varepsilon T_{\mathrm{p}} $ for position subframe in Section \ref{sec:sys_model}-A, \eqref{fim} can be simplified to \eqref{FIM1}.

\end{appendices}
\bibliographystyle{IEEE-unsorted}
\bibliography{refs0308yue}
\end{document}